%% file: main.tex
\documentclass[letterpaper,twocolumn,11pt,floatfix,outputdebug,accepted=2024-04-08]{quantumarticle}
\pdfoutput=1
\usepackage[utf8]{inputenc}
\usepackage[english]{babel}
\usepackage[T1]{fontenc}
\usepackage{amsmath}
\usepackage{amssymb}
\usepackage{hyperref}
\usepackage{tabularx}
\usepackage{mathdots}
\usepackage{braket}
\usepackage{lipsum}
\usepackage{csquotes}
\usepackage[square,numbers,sort&compress]{natbib}

\usepackage{tikz}
\usepackage{tikzit}
\usetikzlibrary{decorations.pathreplacing}
\input{default.tikzstyles}

\newcommand{\symp}[2]{\left\langle #1, #2 \right\rangle}
\DeclareMathOperator{\Tr}{tr}
\DeclareMathOperator{\rank}{rank}

\begin{document}

\title{NoRA: A Tensor Network Ansatz for Volume-Law Entangled Equilibrium States of Highly Connected Hamiltonians}

\author{Val\'erie Bettaque}
\email{vbettaque@brandeis.edu}
\affiliation{Martin A. Fisher School of Physics, Brandeis University, Waltham, MA 02453, USA}
\orcid{0009-0006-9778-8341}

\author{Brian Swingle}
\email{bswingle@brandeis.edu}
\orcid{0000-0003-0334-3108}
\affiliation{Martin A. Fisher School of Physics, Brandeis University, Waltham, MA 02453, USA}

\maketitle

\begin{abstract}
  Motivated by the ground state structure of quantum models with all-to-all interactions such as mean-field quantum spin glass models and the Sachdev-Ye-Kitaev (SYK) model, we propose a tensor network architecture which can accommodate volume law entanglement and a large ground state degeneracy. We call this architecture the non-local renormalization ansatz (NoRA) because it can be viewed as a generalization of MERA, DMERA, and branching MERA networks with the constraints of spatial locality removed. We argue that the architecture is potentially expressive enough to capture the entanglement and complexity of the ground space of the SYK model, thus making it a suitable variational ansatz, but we leave a detailed study of SYK to future work. We further explore the architecture in the special case in which the tensors are random Clifford gates. Here the architecture can be viewed as the encoding map of a random stabilizer code. We introduce a family of codes inspired by the SYK model which can be chosen to have constant rate and linear distance at the cost of some high weight stabilizers. We also comment on potential similarities between this code family and the approximate code formed from the SYK ground space.
\end{abstract}

\vfill

\input{sections/intro}
\input{sections/arch}
\input{sections/clifford}
\input{sections/numerical}
\input{sections/code_struct}
\input{sections/outlook}

\input{glossary}

\bibliographystyle{quantum}
\bibliography{refs}

\onecolumn
\appendix

\input{appendices/entropy_scaling}

\input{appendices/growth_factor}
\input{appendices/phase_space}

\end{document}

%% file: default.tikzstyles
\tikzstyle{rect_dyn}=[fill=white, draw=black, shape=rectangle]
\tikzstyle{rect_m}=[fill=white, draw=black, shape=rectangle, minimum width=1cm, minimum height=0.75cm]
\tikzstyle{rect_l}=[fill=white, draw=black, shape=rectangle, minimum width=2cm, minimum height=0.75cm]
\tikzstyle{brace}=[-, decorate, decoration={{brace}}]

\tikzstyle{default}=[-]
\tikzstyle{dash_default}=[-, dashed, dash pattern=on 0.5mm off 0.5mm]

%% file: sections/intro.tex
\section{Introduction}
\label{sec:intro}

Tensor networks are a powerful tool in the study of geometrically local quantum systems which have proven particularly useful for one-dimensional systems \cite{orus_tensor_networks_complex_2019}. In quantum many-body physics, they first appeared in the guise of \enquote{finitely-correlated states} \cite{fannes_finitely_correlated_states_1992} and were later understood to underlie the functioning of a powerful numerical technique, the density matrix renormalization group (DMRG), which gave unprecedented access to ground states of 1d Hamiltonians~\cite{schollwock_density_matrix_renormalization_2011}. It was understood that DMRG worked because the ground states of interest had limited entanglement and could be effectively compressed to a much smaller space parameterized by so-called matrix product states, a simple kind of 1d tensor network. The use of these tools has since broadened, and there is now a large family of tensor network architectures that are used for both analytical and numerical purposes, both with classical computers and, potentially, quantum computers, with approaches including \cite{haghshenas_variational_power_quantum_2022, kim_robust_entanglement_renormalization_2017, verstraete_matrix_product_density_2004, swingle_renormalization_group_constructions_2016, evenbly_tensor_network_renormalization_2015, czech_tensor_network_quotient_2016, swingle_renormalization_group_circuits_2016, haegeman_rigorous_free_fermion_2018, novikov_exponential_machines_2017, stoudenmire_supervised_learning_quantum_2017, liu_machine_learning_unitary_2019, han_unsupervised_generative_modeling_2018, huggins_quantum_machine_learning_2019, tindall_quantum_physics_connected_2022, searle_thermodynamic_limit_spin_2024}.

In contrast, such network representations have not been much explored for mean-field quantum models which are characterized by all-to-all interactions amongst their degrees of freedom. This is presumably because ground states of such models are expected to be volume-law entangled (e.g.~\cite{liu_quantum_entanglement_sachdev_2018, huang_eigenstate_entanglement_sachdev_2019}), and such a high degree of entanglement is costly to represent using existing tensor networks. In this paper, we address this problem by proposing a class of tensor networks which have the potential to represent the highly entangled ground states of mean-field models.

The networks we consider can be viewed as generalizations of MERA, DMERA, and branching MERA networks where the requirement of spatial locality is removed \cite{vidal_entanglement_renormalization_2007, vidal_class_quantum_many_2008, evenbly_class_highly_entangled_2014, kim_robust_entanglement_renormalization_2017}. As we show below, such networks can accommodate volume law entanglement as is expected for ground states of mean-field models. However, without the imposition of additional structure it is not possible to efficiently contract these networks on a classical computer. Nevertheless, they provide a number of conceptual advantages and can still form the basis for variational quantum algorithms, e.g.~\cite{peruzzo_variational_eigenvalue_solver_2014, mcclean_theory_variational_hybrid_2016}. 

We are particularly motivated to consider these networks in light of the physics of the Sachdev-Ye-Kitaev (SYK) model \cite{sachdev_gapless_spin_fluid_1993, kitaev_simple_model_quantum_2015, polchinski_spectrum_sachdev_ye_2016, maldacena_remarks_sachdev_ye_2016, haldar_variational_wave_functions_2021}. This is a model of all-to-all interacting fermions with a number of unusual features, including an extensive ground state degeneracy and a power-law temperature dependence of the heat capacity at low temperature. Moreover, these curious low energy properties are related to the existence of a dual description in terms of a low-dimensional theory of gravity known as Jackiw-Teitelboim (JT) gravity \cite{kitaev_simple_model_quantum_2015}. It is desirable to better understand the emergence of this gravitational physics, especially for a fixed realization of the couplings, in both the SYK model and beyond. Following earlier ideas relating tensor networks and holography, a small sampling of which is \cite{swingle_entanglement_renormalization_holography_2012, pastawski_holographic_quantum_error_2015, molina-vilaplana_entanglement_tensor_networks_2014, jahn_holography_criticality_matchgate_2019, bhattacharyya_tensor_network_p_2018, bao_toy_models_distilling_2019, jahn_holographic_tensor_network_2021, jahn_tensor_network_models_2022, chen_exact_holographic_tensor_2022}, a tensor network model of SYK may also provide useful information about the emergence of the bulk.

Informed by these properties, we consider a class of networks which can encode an extensively degenerate space of highly entangled ground states. Figure~\ref{fig:arch} illustrates the network architecture, dubbed the non-local renormalization ansatz (NoRA), which should be viewed as a quantum circuit ansatz for the ground space of a suitable class of Hamiltonians. To justify this architecture as a potential model of SYK, we estimate its entanglement and circuit complexity and find qualitative agreement with SYK expectations. In addition to constructing the ground space, the network also provides a skeleton on which we can build a model of excitations~\cite{sewell_thermal_multi_scale_2022}. For an appropriate choice of parameters this model can exhibit a power-law temperature dependence of the thermodynamic entropy (and therefore the heat capacity). These features are the key desiderata underlying our construction, and are discussed in detail in Section~\ref{sec:arch}.

A natural next step would be to explore the NoRA network as a variational ansatz for SYK. This is complicated by two issues: we need to generalize the network structure to fermionic degrees of freedom, and we need to find a way to efficiently contract the network (or use a quantum computer). Given this extra complexity, we have elected to first explore the architecture in a simpler setting where the elementary gates are not variationally chosen but instead are taken to be random Clifford gates. This enables us to study the network properties using the stabilizer formalism~\cite{gottesman_stabilizer_codes_quantum_1997} without needing to explicitly contract the network. Moreover, this setting yields a class of stabilizer codes in which the logical space is identified with the ground state degrees of freedom and the network represents an encoding circuit for the code. We study the stabilizer weights and distance of the resulting codes as a function of the layer depth $D$ and the total system size $N$ (see Figure~\ref{fig:arch}). We find that the network can produce good quantum codes~\cite{calderbank_good_quantum_error_1996}, meaning code families where the distance and number of logical qudits are both proportional to the number of physical qudits. However, these codes are not low-density parity check (LDPC) codes~\cite{gallager_low_density_parity_1962, breuckmann_quantum_low_density_2021} since some of the stabilizers are high weight. We hypothesize that by further fine-tuning the gates, our network architecture could also yield encoding circuits for the recently discovered classes of good quantum LDPC codes~\cite{panteleev_asymptotically_good_quantum_2022, dikstein_locally_testable_codes_2020, dinur_good_quantum_ldpc_2022, lin_good_quantum_ldpc_2022, gu_efficient_decoder_linear_2022, anshu_nlts_hamiltonians_good_2023}.

The rest of this paper is organized as follows: In Section~\ref{sec:arch} we describe the architecture in detail and discuss its key properties. In Section~\ref{sec:clifford} we define a family of random Clifford networks based on our architecture and discuss their interpretation as encoding circuits for stabilizer quantum error correcting codes. In Section~\ref{sec:numerical} we report a numerical study of several different realizations of the architecture falling within the stabilizer code ansatz. We describe in detail how the distance and stabilizer weights of the resulting codes depend on the model parameters. In Section~\ref{sec:code_structure} we discuss a particular thermodynamic limit which is inspired by the structure of SYK. We compare the entanglement and complexity to expectations from holographic calculations and comment on the code properties. Finally, in Section~\ref{sec:outlook} we give an outlook and discuss ongoing and future work.

%% file: sections/arch.tex
\section{Network Architecture}
\label{sec:arch}

Throughout this section we work with general qudits of local dimension $d$. We first describe the general structure of the class of NoRA networks we consider, then we specialize to a particular network structure inspired by scaling and renormalization group (RG) considerations. We analyze both the entanglement and complexity of the scaling-adapted ground state network and discuss an extension to describe excited states. In particular, we show that a natural choice of energy scales in a toy model Hamiltonian can give rise to a power-law temperature dependence of the thermodynamic entropy and heat capacity.

\subsection{General Structure}

\begin{figure}[htb]
    \centering
    \scalebox{0.8}{\input{figures/syk_circuit.tikz}}
    \caption{Basic architecture of the proposed NoRA tensor network ansatz. A code word $\ket{\psi_{\textrm{code}}}$ consisting of $n_0 \equiv k$ (logical) \enquote{ground-state} qudits is embedded  as $\ket{\Psi_{\textrm{phys}}}$ in the (physical) ground space of the $d^N$-dimensional many-body Hilbert space by the means of $L$ layers of some given depth $D$ quantum circuits. For each layer $1 \leq \ell \leq L$ the circuit $D_{\ell}$ acts on the $n_{\ell - 1}$ qudit output from the previous layer and an additional $\Delta n_{\ell}$ new ancillary \enquote{thermal} qudits initialized in state $\ket{0}$. We stress that the layer circuits $D_\ell$ do not have to respect locality structure depicted by the 1d arrangement of qudit lines.}
    \label{fig:arch}
\end{figure}
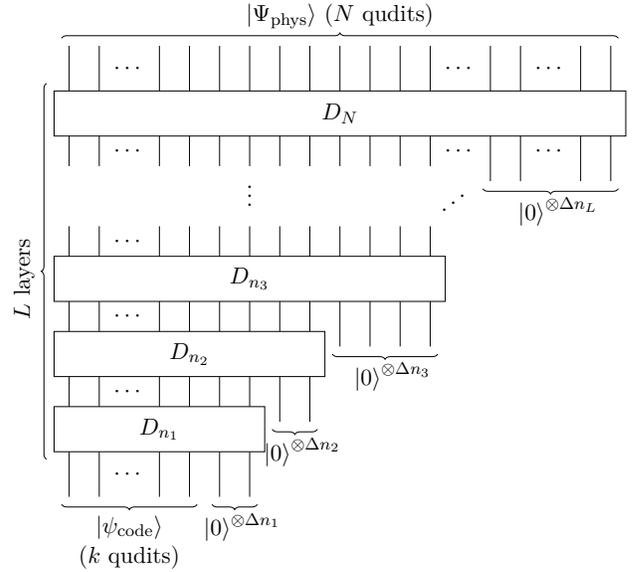

The NoRA network is defined by $L$ layers as in Fig.~\ref{fig:arch}, where we refer to the bottom qudits as \emph{ground state} qudits and the other qudits as \emph{excited state or thermal} qudits. When we set the thermal qudits to some fixed product state, $\ket{0}$, we obtain the \emph{ground state network} as in Fig.~\ref{fig:arch}. This nomenclature is chosen because we can view the network as a variational ansatz for the ground space of a mean-field model. From this point of view, the ground state qudits parameterize a space of states that would be identified with the degenerate ground space of the concrete model of interest.

One way to think about the network is as a ``fine-graining'' circuit moving upwards from the bottom ground state qudits. This is the inverse of a conventional RG transformation since we are adding degrees of freedom. We start with $k$ of these ground state qudits. Then at each layer $\ell$ we add $\Delta n_{\ell}$ thermal qudits in the fixed state $\ket{0}^{\otimes \Delta n_{\ell}}$ and apply a depth $D$ quantum circuit to all the qudits in that layer. This circuit could also be generalized to be time evolution with a suitably normalized all-to-all Hamiltonian for a constant time (proportional to $D$). The next layer takes all the qudits from the previous layer and adds more thermal qudits to generate the hierarchical structure in Fig.~\ref{fig:arch}. The total number of qudits at layer $\ell$ is denoted $n_{\ell}$ and given by
\begin{equation}
    n_{\ell} = k + \sum_{\ell'=1}^{\ell} \Delta n_{\ell'}.
\end{equation}
The total number of qudits is therefore
\begin{equation}
    N \equiv n_L = k + \sum_{\ell=1}^L \Delta n_{\ell}.
\end{equation}

\subsection{Scaling Specialization}

As is, we have described a fairly general architecture. Motivated by scaling and renormalization group considerations, we will primarily consider the special case where $n_{\ell} \sim k + r^{\ell}$, so that the number of thermal qudits is increasing exponentially with each layer up from the bottom. Viewing the top layer as the UV or microscopic degrees of freedom and the bottom layer as the IR or emergent degrees of freedom, moving from the UV to IR (top to bottom) mimics a renormalization group transformation where we remove some fraction of the thermal degrees of freedom at each step. Indeed, borrowing the language of MERA and DMERA and viewing the circuit from top to bottom, the individual layers are like disentanglers that leave behind some decoupled degrees of freedom, the thermal qudits added at that layer. In this scheme, we choose the number of qudits at layer $\ell$ to be
\begin{equation}
    k + r^{\ell} \stackrel{!}{=} n_{\ell} = k + \sum_{\ell'=1}^{\ell} \Delta n_{\ell},
\end{equation}
implying that the number of new thermal qudits for each layer must be
\begin{align}
\begin{split}
    \Delta n_{\ell > 1} &= r^{\ell} - r^{\ell - 1}, \\
    \Delta n_{1} &= r.
\end{split}
\end{align}
For the case of $r=2$, which we primarily consider in this work, this simplifies to approximately $\Delta n_{\ell} = 2^{\ell - 1}$ for all layers $\ell$.

\subsection{Entanglement and Complexity}

We next discuss the entanglement and complexity of the RG-inspired network. There are $\text{O}(N)$ non-trivial bonds in the circuit, of which $N$ bonds connect to the same constant-depth circuit in the last layer. It is therefore straightforward to establish that the network has the potential to encode volume law entanglement for sub-regions of a \enquote{typical} UV state. We also explicitly demonstrate that this is achievable within the Clifford model discussed below in Sections~\ref{sec:clifford}, \ref{sec:numerical}, and \ref{sec:code_structure}.

Turning to the complexity, we take the number of gates in the network as an estimate of the circuit complexity of the UV state, although in general this is only an upper bound. For a layer $\ell$ with $n_{\ell}$ total qubits in it, we apply $D$ rounds of $\lfloor n_{\ell}/q \rfloor$ $q$-qudit gates, so the number of gates of layer $\ell$ is
\begin{equation}
   \text{gates at layer } \ell = D \cdot \lfloor n_{\ell}/q \rfloor.
\end{equation}
Summing this result over all layers and assuming that $q$ divides $n_{\ell}$ without remainder gives a total number of gates equal to
\begin{equation}
\label{eq:circuit_complexity}
    \text{total gates} = \frac{D}{q} \sum_{\ell=1}^L n_{\ell} = \frac{D}{q} \left(L \cdot k + \frac{r^{L+1} - r}{r-1} \right).
\end{equation}
In sections \ref{sec:numerical} and \ref{sec:code_structure} we will cast this result into simpler leading-order expressions that correspond to the respective types of ground space scaling being considered.

\subsection{Extension to Excited States}

Let us conclude this section by extending the ground state network we have so far discussed to the case of excited states. As we have repeatedly emphasized, the discussion so far is general and does not consider a particular physical Hamiltonian. We are simply trying to match certain qualitative features of the entanglement and complexity expected for mean-field models. A structure similar to what we will consider here was recently studied for non-interacting fermions and advocated for as a general approach to approximating thermal states~\cite{sewell_thermal_multi_scale_2022}.

The idea is to introduce a toy Hamiltonian for which the above network is an exact ground state for any choice of state on the $k$ ground state qudits. In other words, the toy Hamiltonian has an exactly degenerate ground space. The Hamiltonian is constructed in a standard way by introducing projectors $P = |0\rangle \langle 0|$ for each thermal qudit and defining corresponding projectors acting on the UV qudits by conjugating these elementary projectors with the network circuit. Let $\tilde{P}_i$ denote the projector for thermal qudit $i$ conjugated by the network circuit. The toy Hamiltonian is
\begin{equation}
\label{eq:stab_hamiltonian}
    H = - \sum_i J_i \tilde{P}_i,
\end{equation}
where $J_i$ are a set of free parameters that determine the energy scale associated with each thermal qudit. Note that -- just like the circuit it encodes -- this Hamiltonian is highly non-local and not necessarily few-body, thus limiting the potential for physical interpretation. The setup is described in more detail in appendix \ref{sec:entropy_scaling}. 

Again motivated by RG considerations, in which the energy scale of excitations decreases by a fixed factor after every RG step (top to bottom), we take the $J_i$ to be equal within a layer and to depend on the layer index $\ell$ as
\begin{equation}
\label{eq:energy_scaling}
    J_{\ell} = \Lambda \cdot e^{-\gamma (L-\ell)}.
\end{equation}
In this way, the UV energy scale is $\Lambda$ and the energy of excitations decreases exponentially with the layer index decreasing towards the IR. The free parameter $\gamma$ controls the rate of decrease.

As computed in appendix \ref{sec:entropy_scaling}, the entropy for the Gibbs ensemble associated to said toy Hamiltonian describing our tensor network ansatz (and for general scaling of $J_{\ell}$) is
\begin{equation}
\label{eq:stab_entropy}
    S = \log\left(d^{k} \cdot (d-1)^{\braket{N-k}}\right) + \sum_{\ell} \Delta n_{\ell} \cdot S(p_{\ell}),
\end{equation}
where we defined a probability, 
\begin{equation}
    p_{\ell} = \frac{d - 1}{e^{\beta J_{\ell}} + d - 1},
\end{equation}
$S(p_{\ell})$ is the classical binary entropy function,
\begin{equation}
    S(p_{\ell}) = - p_{\ell} \cdot \log(p_{\ell}) - (1-p_{\ell}) \cdot \log(1-p_{\ell}),
\end{equation}
and 
\begin{equation}
    \braket{N-k} = \sum_i p_i = \sum_{\ell} \Delta n_{\ell} \, p_{\ell}.
\end{equation}
Note that in the case of qubits ($d=2$), $p_{\ell}$ coincides with the ordinary Fermi-Dirac distribution, in which case $\braket{N-k}$ is analogous to a sum of occupation numbers.

\begin{figure}[htb]
    \centering
    \includegraphics{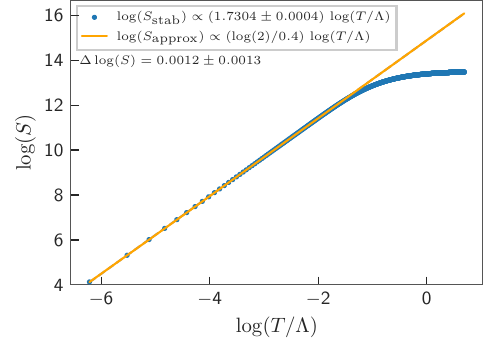}
    \caption{Logarithmic scaling of the exact Gibbs entropy $S_{\textrm{stab}}$ associated to $H$, and the low-temperature approximation $S_{\textrm{approx}}$ for $L=20$, $r=2$, $k=1$, $d=2$, $\Lambda=1$ and $\gamma = 0.4$. Both match almost exactly for our choice of parameters and small $T/\Lambda$, confirming the existence of a scaling law. The same is also true for other choices of $\gamma$ (the only significant free parameter), as seen in figure \ref{fig:entropy_scaling_appendix}.}
    \label{fig:entropy_scaling}
\end{figure}

Plugging in \eqref{eq:energy_scaling} and going to the low-temperature regime (relative to the energy scale $\Lambda$), \eqref{eq:stab_entropy} can be approximated in the continuum limit as
\begin{align}
\label{eq:stab_entropy_approx}
\begin{split}
    & S - k \cdot \log(d) \\ \lessapprox{}& (d-1) (N-k) \cdot \frac{\alpha}{\gamma} \cdot \Gamma\left(\frac{\alpha}{\gamma} + 1\right) (\beta \Lambda)^{-\alpha/\gamma} \\
    \propto{}& (T/\Lambda)^{\alpha/\gamma},
\end{split}
\end{align}
with $N = k + r^L$ and $\alpha = \log(r)$. This together with the specific example depicted in figure \ref{fig:entropy_scaling} confirms that in this limit the entropy does obey a power law. By choosing the parameters $\alpha$ and $\gamma$ suitable, one could even match the precise low-temperature behavior of the SYK heat capacity $C_V$ (which is proportional to $T$) due to $dS = \frac{C_V}{T} dT$:
\begin{equation}
    C_V = T \left( \frac{dS}{dT} \right) \propto (T/\Lambda)^{\alpha/\gamma}.
\end{equation}

\subsection{Summary}

Starting from the general architecture in Figure~\ref{fig:arch}, we introduced the RG-inspired network in which the number of qudits at layer $\ell$ is $k + r^\ell$. In the special case where $k=0$, i.e. a non-degenerate ground space, the number of qudits decreases by a factor from one layer to the next into the IR. This decrease is analogous to a block decimation RG procedure applied to a quantum state. The case of $k\neq 0$ describes a generalization of such an RG procedure. The entanglement entropy of the physical states produced by the RG-inspired network can be volume-law, as expected for mean-field models. We also showed that the ground state network can be extended to provide a model of thermal excitations in which the thermodynamic heat capacity has a power-law temperature dependence at low temperature. These general features are all chosen to match characteristics of the SYK model, which also features a nearly degenerate space of highly entangled ground states and a power-law heat capacity at low temperature.

%% file: figures/syk_circuit.tikz
\begin{tikzpicture}
	\begin{pgfonlayer}{nodelayer}
		\node [style=none] (1) at (-5, -4.5) {};
		\node [style=none] (2) at (-5, -3) {};
		\node [style=none] (3) at (-4, -4.5) {};
		\node [style=none] (4) at (-4, -3) {};
		\node [style=none] (7) at (0, -4.5) {};
		\node [style=none] (8) at (0, -3) {};
		\node [style=none] (11) at (4, 0.5) {};
		\node [style=none] (12) at (4, 2) {};
		\node [style=none] (13) at (-1, -4.5) {};
		\node [style=none] (14) at (-3, -3.75) {$\cdots$};
		\node [style=none] (15) at (-1, -3) {};
		\node [style=none] (19) at (1, -4.5) {};
		\node [style=none] (20) at (1, -3) {};
		\node [style={rect_l}, minimum width=3.5cm] (29) at (-2, -2.25) {$D_{n_1}$};
		\node [style=none] (42) at (-3, -5.5) {$\ket{\psi_{\text{code}}}$};
		\node [style=none] (43) at (0.75, -5.5) {$\ket{0}^{\otimes \Delta n_1}$};
		\node [style=none] (44) at (2, -2) {};
		\node [style=none] (45) at (2, -0.5) {};
		\node [style=none] (46) at (3, -2) {};
		\node [style=none] (47) at (3, -0.5) {};
		\node [style={rect_l}, minimum width=4.5cm] (48) at (-1, 0.25) {$D_{n_2}$};
		\node [style=none] (49) at (5, 2) {};
		\node [style=none] (50) at (5, 0.5) {};
		\node [style=none] (51) at (-5, -1.5) {};
		\node [style=none] (52) at (-5, -0.5) {};
		\node [style=none] (53) at (-4, -1.5) {};
		\node [style=none] (54) at (-4, -0.5) {};
		\node [style=none] (59) at (1, -1.5) {};
		\node [style=none] (60) at (1, -0.5) {};
		\node [style=none] (61) at (0, -1.5) {};
		\node [style=none] (62) at (0, -0.5) {};
		\node [style=none] (63) at (-1, -1.5) {};
		\node [style=none] (64) at (-1, -0.5) {};
		\node [style=none] (65) at (-3, -1) {$\cdots$};
		\node [style=none] (66) at (-2, -4.5) {};
		\node [style=none] (67) at (-2, -3) {};
		\node [style=none] (70) at (-2, -1.5) {};
		\node [style=none] (71) at (-2, -0.5) {};
		\node [style={rect_l}, minimum width=9.5cm] (72) at (4, 8.25) {$D_{N}$};
		\node [style=none] (79) at (1, 5.75) {$\vdots$};
		\node [style=none] (80) at (-5, 1) {};
		\node [style=none] (81) at (-4, 2) {};
		\node [style=none] (82) at (-2, 1) {};
		\node [style=none] (83) at (-5, 2) {};
		\node [style=none] (85) at (-4, 1) {};
		\node [style=none] (86) at (-2, 2) {};
		\node [style=none] (87) at (-1, 2) {};
		\node [style=none] (88) at (-1, 1) {};
		\node [style=none] (89) at (0, 2) {};
		\node [style=none] (90) at (0, 1) {};
		\node [style=none] (91) at (1, 2) {};
		\node [style=none] (92) at (1, 1) {};
		\node [style=none] (93) at (-3, 1.5) {$\cdots$};
		\node [style=none] (94) at (2, 1) {};
		\node [style=none] (95) at (3, 1) {};
		\node [style=none] (96) at (2, 2) {};
		\node [style=none] (97) at (3, 2) {};
		\node [style=none] (98) at (-5, 6.5) {};
		\node [style=none] (99) at (-4, 7.5) {};
		\node [style=none] (100) at (-2, 6.5) {};
		\node [style=none] (101) at (-5, 7.5) {};
		\node [style=none] (102) at (-4, 6.5) {};
		\node [style=none] (103) at (-2, 7.5) {};
		\node [style=none] (104) at (-1, 7.5) {};
		\node [style=none] (105) at (-1, 6.5) {};
		\node [style=none] (106) at (0, 7.5) {};
		\node [style=none] (107) at (0, 6.5) {};
		\node [style=none] (108) at (1, 7.5) {};
		\node [style=none] (109) at (1, 6.5) {};
		\node [style=none] (110) at (-3, 7) {$\cdots$};
		\node [style=none] (111) at (2, 6.5) {};
		\node [style=none] (112) at (3, 6.5) {};
		\node [style=none] (113) at (2, 7.5) {};
		\node [style=none] (114) at (3, 7.5) {};
		\node [style=none] (117) at (-5, 9) {};
		\node [style=none] (118) at (-4, 10.5) {};
		\node [style=none] (119) at (-2, 9) {};
		\node [style=none] (120) at (-5, 10.5) {};
		\node [style=none] (121) at (-4, 9) {};
		\node [style=none] (122) at (-2, 10.5) {};
		\node [style=none] (123) at (-1, 10.5) {};
		\node [style=none] (124) at (-1, 9) {};
		\node [style=none] (125) at (0, 10.5) {};
		\node [style=none] (126) at (0, 9) {};
		\node [style=none] (127) at (1, 10.5) {};
		\node [style=none] (128) at (1, 9) {};
		\node [style=none] (129) at (-3, 9.75) {$\cdots$};
		\node [style=none] (130) at (2, 9) {};
		\node [style=none] (131) at (3, 9) {};
		\node [style=none] (132) at (2, 10.5) {};
		\node [style=none] (133) at (3, 10.5) {};
		\node [style=none] (135) at (4, 9) {};
		\node [style=none] (136) at (5, 9) {};
		\node [style=none] (137) at (4, 10.5) {};
		\node [style=none] (138) at (5, 10.5) {};
		\node [style=none] (140) at (-0.75, -4.75) {};
		\node [style=none] (141) at (-5.25, -4.75) {};
		\node [style=none] (142) at (1.25, -4.75) {};
		\node [style=none] (143) at (-0.25, -4.75) {};
		\node [style=none] (144) at (-5.25, 10.75) {};
		\node [style=none] (145) at (13.25, 10.75) {};
		\node [style=none] (146) at (-5.75, -3.25) {};
		\node [style=none] (147) at (-5.75, 9.25) {};
		\node [style=none, rotate=90] (148) at (-6.5, 2.75) {$L$ layers};
		\node [style=none] (149) at (4, 11.5) {$\ket{\Psi_{\text{phys}}}$ ($N$ qudits)};
		\node [style={rect_l}, minimum width=6.5cm] (150) at (1, 2.75) {$D_{n_3}$};
		\node [style=none] (151) at (6, 0.5) {};
		\node [style=none] (152) at (6, 2) {};
		\node [style=none] (153) at (7, 2) {};
		\node [style=none] (154) at (7, 0.5) {};
		\node [style=none] (155) at (-5, 3.5) {};
		\node [style=none] (156) at (-4, 4.5) {};
		\node [style=none] (157) at (-2, 3.5) {};
		\node [style=none] (158) at (-5, 4.5) {};
		\node [style=none] (159) at (-4, 3.5) {};
		\node [style=none] (160) at (-2, 4.5) {};
		\node [style=none] (161) at (-1, 4.5) {};
		\node [style=none] (162) at (-1, 3.5) {};
		\node [style=none] (163) at (0, 4.5) {};
		\node [style=none] (164) at (0, 3.5) {};
		\node [style=none] (165) at (1, 4.5) {};
		\node [style=none] (166) at (1, 3.5) {};
		\node [style=none] (167) at (-3, 4) {$\cdots$};
		\node [style=none] (168) at (2, 3.5) {};
		\node [style=none] (169) at (3, 3.5) {};
		\node [style=none] (170) at (2, 4.5) {};
		\node [style=none] (171) at (3, 4.5) {};
		\node [style=none] (172) at (4, 4.5) {};
		\node [style=none] (173) at (4, 3.5) {};
		\node [style=none] (174) at (5, 4.5) {};
		\node [style=none] (175) at (5, 3.5) {};
		\node [style=none] (176) at (6, 3.5) {};
		\node [style=none] (177) at (7, 3.5) {};
		\node [style=none] (178) at (6, 4.5) {};
		\node [style=none] (179) at (7, 4.5) {};
		\node [style=none] (180) at (2.75, -3) {$\ket{0}^{\otimes \Delta n_2}$};
		\node [style=none] (181) at (3.25, -2.25) {};
		\node [style=none] (182) at (1.75, -2.25) {};
		\node [style=none] (183) at (5.75, -0.5) {$\ket{0}^{\otimes \Delta n_3}$};
		\node [style=none] (184) at (7.25, 0.25) {};
		\node [style=none] (185) at (3.75, 0.25) {};
		\node [style=none] (186) at (4, 7.5) {};
		\node [style=none] (187) at (4, 6.5) {};
		\node [style=none] (188) at (5, 7.5) {};
		\node [style=none] (189) at (5, 6.5) {};
		\node [style=none] (190) at (6, 6.5) {};
		\node [style=none] (191) at (7, 6.5) {};
		\node [style=none] (192) at (6, 7.5) {};
		\node [style=none] (193) at (7, 7.5) {};
		\node [style=none] (194) at (6, 10.5) {};
		\node [style=none] (195) at (6, 9) {};
		\node [style=none] (196) at (7, 10.5) {};
		\node [style=none] (197) at (7, 9) {};
		\node [style=none] (198) at (9, 9) {};
		\node [style=none] (199) at (10, 9) {};
		\node [style=none] (200) at (9, 10.5) {};
		\node [style=none] (201) at (10, 10.5) {};
		\node [style=none] (202) at (11, 9.75) {$\cdots$};
		\node [style=none] (203) at (8, 9.75) {$\cdots$};
		\node [style=none] (204) at (12, 9) {};
		\node [style=none] (205) at (13, 9) {};
		\node [style=none] (206) at (12, 10.5) {};
		\node [style=none] (207) at (13, 10.5) {};
		\node [style=none] (208) at (12, 6) {};
		\node [style=none] (209) at (13, 6) {};
		\node [style=none] (210) at (12, 7.5) {};
		\node [style=none] (211) at (13, 7.5) {};
		\node [style=none] (212) at (9, 6) {};
		\node [style=none] (213) at (10, 6) {};
		\node [style=none] (214) at (9, 7.5) {};
		\node [style=none] (215) at (10, 7.5) {};
		\node [style=none] (216) at (11, 7) {$\cdots$};
		\node [style=none] (217) at (8, 7) {$\cdots$};
		\node [style=none] (218) at (11.25, 5) {$\ket{0}^{\otimes \Delta n_L}$};
		\node [style=none] (219) at (13.25, 5.75) {};
		\node [style=none] (220) at (8.75, 5.75) {};
		\node [style=none] (221) at (7.75, 5.5) {$\iddots$};
		\node [style=none] (222) at (2, 18) {};
		\node [style=none] (223) at (-3, -6.5) {($k$ qudits)};
	\end{pgfonlayer}
	\begin{pgfonlayer}{edgelayer}
		\draw [style=default] (1.center) to (2.center);
		\draw [style=default] (3.center) to (4.center);
		\draw [style=default] (7.center) to (8.center);
		\draw [style=default] (11.center) to (12.center);
		\draw [style=default] (13.center) to (15.center);
		\draw [style=default] (19.center) to (20.center);
		\draw [style=default] (44.center) to (45.center);
		\draw [style=default] (46.center) to (47.center);
		\draw [style=default] (50.center) to (49.center);
		\draw [style=default] (51.center) to (52.center);
		\draw [style=default] (53.center) to (54.center);
		\draw [style=default] (60.center) to (59.center);
		\draw [style=default] (62.center) to (61.center);
		\draw [style=default] (64.center) to (63.center);
		\draw [style=default] (67.center) to (66.center);
		\draw [style=default] (71.center) to (70.center);
		\draw [style=default] (83.center) to (80.center);
		\draw [style=default] (85.center) to (81.center);
		\draw [style=default] (92.center) to (91.center);
		\draw [style=default] (90.center) to (89.center);
		\draw [style=default] (88.center) to (87.center);
		\draw [style=default] (82.center) to (86.center);
		\draw [style=default] (94.center) to (96.center);
		\draw [style=default] (95.center) to (97.center);
		\draw [style=default] (101.center) to (98.center);
		\draw [style=default] (102.center) to (99.center);
		\draw [style=default] (109.center) to (108.center);
		\draw [style=default] (107.center) to (106.center);
		\draw [style=default] (105.center) to (104.center);
		\draw [style=default] (100.center) to (103.center);
		\draw [style=default] (111.center) to (113.center);
		\draw [style=default] (112.center) to (114.center);
		\draw [style=default] (120.center) to (117.center);
		\draw [style=default] (121.center) to (118.center);
		\draw [style=default] (128.center) to (127.center);
		\draw [style=default] (126.center) to (125.center);
		\draw [style=default] (124.center) to (123.center);
		\draw [style=default] (119.center) to (122.center);
		\draw [style=default] (130.center) to (132.center);
		\draw [style=default] (131.center) to (133.center);
		\draw [style=default] (135.center) to (137.center);
		\draw [style=default] (136.center) to (138.center);
		\draw [style=brace] (140.center) to (141.center);
		\draw [style=brace] (142.center) to (143.center);
		\draw [style=brace] (144.center) to (145.center);
		\draw [style=brace] (146.center) to (147.center);
		\draw [style=default] (151.center) to (152.center);
		\draw [style=default] (154.center) to (153.center);
		\draw [style=default] (158.center) to (155.center);
		\draw [style=default] (159.center) to (156.center);
		\draw [style=default] (166.center) to (165.center);
		\draw [style=default] (164.center) to (163.center);
		\draw [style=default] (162.center) to (161.center);
		\draw [style=default] (157.center) to (160.center);
		\draw [style=default] (168.center) to (170.center);
		\draw [style=default] (169.center) to (171.center);
		\draw [style=default] (175.center) to (174.center);
		\draw [style=default] (173.center) to (172.center);
		\draw [style=default] (176.center) to (178.center);
		\draw [style=default] (177.center) to (179.center);
		\draw [style=brace] (181.center) to (182.center);
		\draw [style=brace] (184.center) to (185.center);
		\draw [style=default] (189.center) to (188.center);
		\draw [style=default] (187.center) to (186.center);
		\draw [style=default] (190.center) to (192.center);
		\draw [style=default] (191.center) to (193.center);
		\draw [style=default] (197.center) to (196.center);
		\draw [style=default] (195.center) to (194.center);
		\draw [style=default] (198.center) to (200.center);
		\draw [style=default] (199.center) to (201.center);
		\draw [style=default] (204.center) to (206.center);
		\draw [style=default] (205.center) to (207.center);
		\draw [style=default] (208.center) to (210.center);
		\draw [style=default] (209.center) to (211.center);
		\draw [style=default] (212.center) to (214.center);
		\draw [style=default] (213.center) to (215.center);
		\draw [style=brace] (219.center) to (220.center);
	\end{pgfonlayer}
\end{tikzpicture}

%% file: sections/clifford.tex
\section{Clifford Ansatz}
\label{sec:clifford}

Having laid out the scaling-inspired architecture in the previous section and shown that it can capture some expected features of mean-field models, especially the SYK model, we now consider a concrete version of the network built from Clifford gates. We would also like to use the network as a variational ansatz to study physical mean-field models, but for the reasons outlined in the introduction, in this paper we focus on the Clifford model as an example where we can also classically simulate the network properties. A review of the Clifford group and how it can be implemented is provided in appendix \ref{sec:phase_space}. 

If the circuits in Figure~\ref{fig:arch} are composed of Clifford gates, then the network can be interpreted as an encoding circuit for a stabilizer quantum error correcting code~\cite{gottesman_stabilizer_codes_quantum_1997}. The ground state qudits then correspond to the logical qudits of the code. We focus in particular on the distance of the code and the weight of the stabilizers, as they provide a good heuristic for probing the entanglement structure and give us a glimpse at the network's potential as an error-correcting code. The purpose of this section is to review this error correction interpretation and setup the subsequent calculations in Sections~\ref{sec:numerical} and \ref{sec:code_structure}.

\subsection{The Clifford Group}

Let us briefly recall the motivation for Clifford circuits. In general, simulating quantum circuits on a classical computer architecture becomes difficult with increasing number of qudits due to the exponential scaling of the Hilbert space dimension with the number of qudits. However, we can still compute certain quantities efficiently on classical computers by restricting ourselves to a subgroup of the full unitary group that only scales linearly in the number of qudits ~\cite{gottesman_heisenberg_representation_quantum_1998, aaronson_improved_simulation_stabilizer_2004}. This group is called the \emph{Clifford group} and is defined as the subgroup of unitary operators that map Pauli strings to Pauli strings~\cite{gottesman_heisenberg_representation_quantum_1998}. Elements of the Clifford group can then be represented as \emph{Clifford circuits}, which are circuits composed of successive (elementary) Clifford gates acting on a bounded number of qudits at a time. An example of such a Clifford circuit is depicted in Figure \ref{fig:layer_circuit} in the context of random scrambling.

The Clifford group has a variety of applications in quantum information. For example, in the context of generating random states, the Clifford group is useful because it forms a $k$-design of the Haar measure of random unitaries. This means that quantities averaged over random choices of gates/states only start to differ between Clifford and Haar in probability moments higher than $k$, where we have $k=1$ for all possible qudit dimensions $d$, $k=2$ for all $d$ that are powers of primes, and $k=3$ when said base prime is 2. \cite{zhu_multiqubit_clifford_groups_2017, graydon_clifford_groups_are_2021}.

The reason the Clifford group for $N$ qudits with local dimension $d$ can be efficiently simulated resides in the fact that it is a projective representation of the symplectic group $\text{Sp}(2N, \mathbb{F}_d)$, where $2N$ is the vector space dimension the group acts on and $\mathbb{F}_d$ is the (unique) finite field with $d$\footnote{In this case $d$ must be the power of a prime.} elements. As mentioned before, the space of Pauli strings therefore scales linearly, with operators mapping between them being represented (up to a phase) by $2N \times 2N$ symplectic matrices over $\mathbb{F}_d$. Sampling Cliffords therefore can be achieved by sampling symplectic matrices, for which efficient algorithms exist \cite{koenig_how_efficiently_select_2014}.  A more detailed description of this framework is provided in Appendix \ref{sec:phase_space}.

\subsection{Random Layer Circuits}

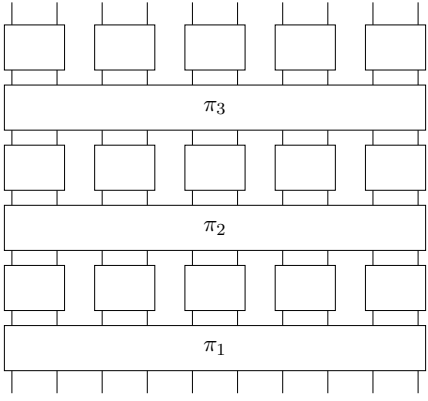
\begin{figure}[hbt]
    \centering
    \scalebox{0.8}{\input{figures/wall_circuit.tikz}}
    \caption{An example of a layer circuit acting on 10 qudits with $q=2$ and depth $D=3$. Each unmarked gate represents a randomly sampled Clifford element acting on two qudits, while the gates $\pi_m$ for $m=1,2,3$ are random permutations of the qudits. While such a circuit does not exhibit a causal cone, this non-locality of interactions is expected from mean-field models.}
    \label{fig:layer_circuit}
\end{figure}

To define a precise model based on our architecture, we have to make an explicit choice for the depth $D$ circuits $D_{\ell}$ that are applied at each layer $\ell$. Inspired by SYK, our approach is to apply $n_{\ell} / q$ randomly sampled Clifford gates\footnote{In general $q$ does not have to divide $n_{\ell}$ without remainder, in which case one can simply leave $n_{\ell} \mod{q}$ qudits unchanged at each sub-layer. In our computations we always chose our parameters such that this is not necessary. } to randomly chosen non-intersecting sets of $q$ qudits for each sub-layer $1 \leq m \leq D$ of the total layer circuit. Such a Clifford circuit is depicted in figure \ref{fig:layer_circuit} for $q=2$. Heuristically, this ansatz can be interpreted as a Trotterization of the SYK Hamiltonian, although with qudits instead of Majorana fermions \footnote{In an actual variational calculation with the SYK model, we might expect that the layer circuits are unitarites generated by SYK-Hamiltonian-like operators (although not necessarily the SYK Hamiltonian itself). The choice of random Clifford layers is thus loosely inspired by our expectations for the SYK model.} .

With that we can then view the resulting network as an encoding circuit for a quantum stabilizer code. The $k$ ground state qudits are the logical qudits and the $N$ UV qudits are the physical qudits. We now briefly review stabilizer codes and the important notion of distance, which captures aspects of the entanglement structure discussed above in Section~\ref{sec:arch}.

\subsection{Stabilizer Codes}
\label{sec:stab}

A $[[N, k, \delta]]$ \emph{stabilizer code} that encodes $k$ logical qudits into $N$ physical qudits with distance $\delta$ is defined in terms of a \emph{stabilizer group} $S$, which is an abelian subgroup of the (generalized) Pauli group $P_d(N)$ i.e.\ the group generated by all possible $N$-element tensor products of ordinary Pauli operators ($d = 2$)
\begin{equation}
    X = \begin{pmatrix}
        0 & 1 \\ 1 & 0
    \end{pmatrix}, \,
    Y = \begin{pmatrix}
        0 & -i \\ i & 0
    \end{pmatrix}, \,
    Z = \begin{pmatrix}
        1 & 0 \\ 0 & -1
    \end{pmatrix},
\end{equation}
or their higher-dimensional counterparts ($d > 2$), which are defined in appendix \ref{sec:weyl_rep} \cite{gottesman_stabilizer_codes_quantum_1997}. The stabilizer group must therefore be generated by $N - k$ independent and commuting elements of $P_d(N)$. A \emph{code word} then is a state vector $\ket{\psi} \in \mathbb{C}^{d^N}$ that satisfies $s \ket{\psi} = \ket{\psi}$ for all $s \in S$. The space spanned by all possible code words is called the \emph{code space} and has dimension $d^k$ due to the rank of the group being $N-k$. The operators mapping logical states to other logical states are called \emph{logical operators} and must therefore commute with all elements of the stabilizer group and hence form the centralizer of the stabilizer group in in $P_d(N)$.

\subsubsection{Decoupling \& Code Distance}

\begin{figure}[htb]
    \centering
    \scalebox{1.1}{\input{figures/decoupling.tikz}}
    \caption{Circuit representation of state in which the code space is maximally entangled with a reference $R$. Here $U_{MN}$ is a unitary that takes states of the form $\ket{\psi_{\textrm{anc}}}_M \otimes \ket{\psi_{\textrm{code}}}_N$ and maps $\ket{\psi_{\textrm{code}}}$ to the code space of the chosen stabilizer code. $\ket{\psi_{\textrm{anc}}}$ is the all $0$ state of the ancillary qubits. If a region $A$ has zero mutual information with $R$, then it has no access to the encoded information. The code distance $\delta$ is the biggest integer such that all regions $A$ of size $|A| < \delta$ have $I(A,R)=0$.}
    \label{fig:decoupling}
\end{figure}
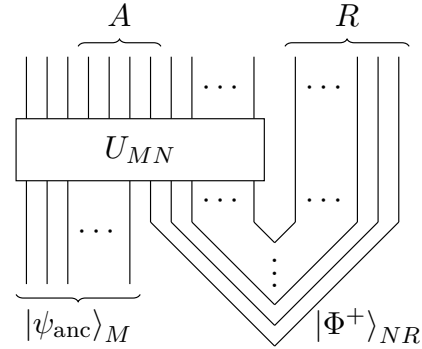

The code distance is a measure of how robust the code is to errors on the physical qubits. Determining the distance for a stabilizer code is in general a computationally intensive problem due to the potential for complex patterns of entanglement. We use an \emph{adversarial approach}, which is based on analyzing the mutual information
\begin{equation}
\label{eq:mut_inf}
    I(A, R) = S(A) + S(R) - S(AR)
\end{equation}
between all possible subsystems $A$ of the physical qudits and some external reference $R$ which is maximally entangled with the code space. A depiction of the setup can be found in figure \ref{fig:decoupling}.

Because $R$ is maximally entangled with the code space, it is effectively tracking the encoded information. Therefore the question is how much of the system does an adversary need access to in order to be correlated with $R$ and thus have (at least partial) access to the encoded information. This correlation can be detected using the aforementioned mutual information \eqref{eq:mut_inf}, which becomes non-zero in such a case.
The code distance $\delta$ is the biggest integer such that all regions $A$ with $|A|<\delta$ have $I(A,R)=0$. 

Implementing this approach as an algorithm is time-consuming though, since iterating through all possible choices for $A$ is combinatorically intensive. A way to simplify the procedure at the cost of only getting an upper bound approximation for the code distance is by randomly sampling choices for $A$ and determining the largest one which has vanishing mutual information. This Monte Carlo approach is the method we use.

\subsubsection{Stabilizer Weights}

It is also interesting to ask about the weights of the stabilizers. The weight of a Pauli string is defined as the number of elements of $P_d(1)$ in the tensor product representation of the operator that are not proportional to the identity operator $I$. Since $P_d(1)$ contains $d^2$ elements, there are $d^2 - 1$ such nontrivial operators.  If the stabilizer group has a generating set containing only Pauli strings of bounded weight, then we say the code has constant weight. The code space can always be obtained as the ground space of a Hamiltonian built from a generating set of the stabilizer group, and if the code has constant weight then there is such a Hamiltonian which contains only terms acting on a bounded number of qudits at a time.

\subsection{Summary}

Here we reviewed the notion of a stabilizer code and defined the random Clifford gate version of our architecture. In the following two sections, \ref{sec:numerical} and \ref{sec:code_structure}, we consider random stabilizer codes built from random Clifford layers inserted in the RG-inspired architecture (Figure~\ref{fig:arch} and Section~\ref{sec:arch}). We investigate the distance and stabilizer weights both numerically and via analytic arguments. We verify that these codes can be highly entangled, for example, with a distance proportional to $N$. We also study the distribution of stabilizer weights and show that some stabilizers do have high weight proportional to $N$. As such, they are not constant weight codes in general.

%% file: figures/wall_circuit.tikz
\begin{tikzpicture}
	\begin{pgfonlayer}{nodelayer}
		\node [style=none] (15) at (-2.25, 1.25) {};
		\node [style=none] (16) at (-0.75, 1.25) {};
		\node [style=none] (17) at (-2.25, 1.75) {};
		\node [style=none] (18) at (-0.75, 1.75) {};
		\node [style={rect_m}] (19) at (1.5, 2.5) {};
		\node [style=none] (20) at (0.75, 1.25) {};
		\node [style=none] (21) at (2.25, 1.25) {};
		\node [style=none] (22) at (0.75, 1.75) {};
		\node [style=none] (23) at (2.25, 1.75) {};
		\node [style=none] (24) at (3.75, 1.25) {};
		\node [style=none] (25) at (3.75, 1.75) {};
		\node [style=none] (26) at (-3.75, 1.25) {};
		\node [style=none] (27) at (-3.75, 1.75) {};
		\node [style={rect_m}] (28) at (-1.5, 2.5) {};
		\node [style={rect_m}] (29) at (4.5, 2.5) {};
		\node [style={rect_m}] (30) at (-4.5, 2.5) {};
		\node [style=none] (31) at (-2.25, 3.25) {};
		\node [style=none] (32) at (-0.75, 3.25) {};
		\node [style=none] (33) at (-2.25, 3.75) {};
		\node [style=none] (34) at (-0.75, 3.75) {};
		\node [style=none] (36) at (0.75, 3.25) {};
		\node [style=none] (37) at (2.25, 3.25) {};
		\node [style=none] (38) at (0.75, 3.75) {};
		\node [style=none] (39) at (2.25, 3.75) {};
		\node [style=none] (40) at (3.75, 3.25) {};
		\node [style=none] (41) at (3.75, 3.75) {};
		\node [style=none] (42) at (-3.75, 3.25) {};
		\node [style=none] (43) at (-3.75, 3.75) {};
		\node [style=none] (46) at (-2.25, 5.25) {};
		\node [style=none] (47) at (-0.75, 5.25) {};
		\node [style=none] (48) at (-2.25, 5.75) {};
		\node [style=none] (49) at (-0.75, 5.75) {};
		\node [style={rect_m}] (50) at (1.5, 6.5) {};
		\node [style=none] (51) at (0.75, 5.25) {};
		\node [style=none] (52) at (2.25, 5.25) {};
		\node [style=none] (53) at (0.75, 5.75) {};
		\node [style=none] (54) at (2.25, 5.75) {};
		\node [style=none] (55) at (3.75, 5.25) {};
		\node [style=none] (56) at (3.75, 5.75) {};
		\node [style=none] (57) at (-3.75, 5.25) {};
		\node [style=none] (58) at (-3.75, 5.75) {};
		\node [style={rect_m}] (59) at (-1.5, 6.5) {};
		\node [style={rect_m}] (60) at (4.5, 6.5) {};
		\node [style={rect_m}] (61) at (-4.5, 6.5) {};
		\node [style=none] (62) at (-2.25, 7.25) {};
		\node [style=none] (63) at (-0.75, 7.25) {};
		\node [style=none] (64) at (-2.25, 8) {};
		\node [style=none] (65) at (-0.75, 8) {};
		\node [style=none] (66) at (0.75, 7.25) {};
		\node [style=none] (67) at (2.25, 7.25) {};
		\node [style=none] (68) at (0.75, 8) {};
		\node [style=none] (69) at (2.25, 8) {};
		\node [style=none] (70) at (3.75, 7.25) {};
		\node [style=none] (71) at (3.75, 8) {};
		\node [style=none] (72) at (-3.75, 7.25) {};
		\node [style=none] (73) at (-3.75, 8) {};
		\node [style=none] (86) at (-5.25, 1.25) {};
		\node [style=none] (87) at (-5.25, 1.75) {};
		\node [style=none] (88) at (-6.75, 1.25) {};
		\node [style=none] (89) at (-6.75, 1.75) {};
		\node [style=none] (90) at (-5.25, 3.25) {};
		\node [style=none] (91) at (-5.25, 3.75) {};
		\node [style=none] (92) at (-6.75, 3.25) {};
		\node [style=none] (93) at (-6.75, 3.75) {};
		\node [style=none] (94) at (-5.25, 5.25) {};
		\node [style=none] (95) at (-5.25, 5.75) {};
		\node [style=none] (96) at (-6.75, 5.25) {};
		\node [style=none] (97) at (-6.75, 5.75) {};
		\node [style=none] (98) at (-5.25, 7.25) {};
		\node [style=none] (99) at (-5.25, 8) {};
		\node [style=none] (100) at (-6.75, 7.25) {};
		\node [style=none] (101) at (-6.75, 8) {};
		\node [style=none] (104) at (5.25, 1.25) {};
		\node [style=none] (105) at (5.25, 1.75) {};
		\node [style=none] (108) at (5.25, 3.25) {};
		\node [style=none] (109) at (5.25, 3.75) {};
		\node [style=none] (112) at (5.25, 5.25) {};
		\node [style=none] (113) at (5.25, 5.75) {};
		\node [style=none] (116) at (5.25, 7.25) {};
		\node [style=none] (117) at (5.25, 8) {};
		\node [style={rect_m}] (120) at (-7.5, 2.5) {};
		\node [style={rect_m}] (121) at (-7.5, 6.5) {};
		\node [style=none] (126) at (-8.25, 1.25) {};
		\node [style=none] (127) at (-8.25, 1.75) {};
		\node [style=none] (133) at (-2.25, -2.75) {};
		\node [style=none] (134) at (-0.75, -2.75) {};
		\node [style=none] (135) at (-2.25, -2.25) {};
		\node [style=none] (136) at (-0.75, -2.25) {};
		\node [style={rect_m}] (137) at (1.5, -1.5) {};
		\node [style=none] (138) at (0.75, -2.75) {};
		\node [style=none] (139) at (2.25, -2.75) {};
		\node [style=none] (140) at (0.75, -2.25) {};
		\node [style=none] (141) at (2.25, -2.25) {};
		\node [style=none] (142) at (3.75, -2.75) {};
		\node [style=none] (143) at (3.75, -2.25) {};
		\node [style=none] (144) at (-3.75, -2.75) {};
		\node [style=none] (145) at (-3.75, -2.25) {};
		\node [style={rect_m}] (146) at (-1.5, -1.5) {};
		\node [style={rect_m}] (147) at (4.5, -1.5) {};
		\node [style={rect_m}] (148) at (-4.5, -1.5) {};
		\node [style=none] (149) at (-2.25, -0.75) {};
		\node [style=none] (150) at (-0.75, -0.75) {};
		\node [style=none] (151) at (-2.25, -0.25) {};
		\node [style=none] (152) at (-0.75, -0.25) {};
		\node [style=none] (153) at (0.75, -0.75) {};
		\node [style=none] (154) at (2.25, -0.75) {};
		\node [style=none] (155) at (0.75, -0.25) {};
		\node [style=none] (156) at (2.25, -0.25) {};
		\node [style=none] (157) at (3.75, -0.75) {};
		\node [style=none] (158) at (3.75, -0.25) {};
		\node [style=none] (159) at (-3.75, -0.75) {};
		\node [style=none] (160) at (-3.75, -0.25) {};
		\node [style=none] (161) at (-5.25, -2.75) {};
		\node [style=none] (162) at (-5.25, -2.25) {};
		\node [style=none] (163) at (-6.75, -2.75) {};
		\node [style=none] (164) at (-6.75, -2.25) {};
		\node [style=none] (165) at (-5.25, -0.75) {};
		\node [style=none] (166) at (-5.25, -0.25) {};
		\node [style=none] (167) at (-6.75, -0.75) {};
		\node [style=none] (168) at (-6.75, -0.25) {};
		\node [style=none] (171) at (5.25, -2.75) {};
		\node [style=none] (172) at (5.25, -2.25) {};
		\node [style=none] (175) at (5.25, -0.75) {};
		\node [style=none] (176) at (5.25, -0.25) {};
		\node [style={rect_m}] (178) at (-7.5, -1.5) {};
		\node [style=none] (179) at (-8.25, 3.25) {};
		\node [style=none] (180) at (-8.25, 3.75) {};
		\node [style=none] (181) at (-8.25, 5.25) {};
		\node [style=none] (182) at (-8.25, 5.75) {};
		\node [style=none] (183) at (-8.25, 7.25) {};
		\node [style=none] (184) at (-8.25, 8) {};
		\node [style=none] (185) at (-8.25, -2.75) {};
		\node [style=none] (186) at (-8.25, -2.25) {};
		\node [style=none] (187) at (-8.25, -0.75) {};
		\node [style=none] (188) at (-8.25, -0.25) {};
		\node [style={rect_l}, minimum width=7cm] (189) at (-1.5, 0.5) {$\pi_2$};
		\node [style={rect_l}, minimum width=7cm] (190) at (-1.5, 4.5) {$\pi_3$};
		\node [style={rect_l}, minimum width=7cm] (191) at (-1.5, -3.5) {$\pi_1$};
		\node [style=none] (192) at (-2.25, -5) {};
		\node [style=none] (193) at (-0.75, -5) {};
		\node [style=none] (194) at (-2.25, -4.25) {};
		\node [style=none] (195) at (-0.75, -4.25) {};
		\node [style=none] (196) at (0.75, -5) {};
		\node [style=none] (197) at (2.25, -5) {};
		\node [style=none] (198) at (0.75, -4.25) {};
		\node [style=none] (199) at (2.25, -4.25) {};
		\node [style=none] (200) at (3.75, -5) {};
		\node [style=none] (201) at (3.75, -4.25) {};
		\node [style=none] (202) at (-3.75, -5) {};
		\node [style=none] (203) at (-3.75, -4.25) {};
		\node [style=none] (204) at (-5.25, -5) {};
		\node [style=none] (205) at (-5.25, -4.25) {};
		\node [style=none] (206) at (-6.75, -5) {};
		\node [style=none] (207) at (-6.75, -4.25) {};
		\node [style=none] (208) at (5.25, -5) {};
		\node [style=none] (209) at (5.25, -4.25) {};
		\node [style=none] (210) at (-8.25, -5) {};
		\node [style=none] (211) at (-8.25, -4.25) {};
	\end{pgfonlayer}
	\begin{pgfonlayer}{edgelayer}
		\draw [style=default] (15.center) to (17.center);
		\draw [style=default] (16.center) to (18.center);
		\draw [style=default] (20.center) to (22.center);
		\draw [style=default] (21.center) to (23.center);
		\draw [style=default] (24.center) to (25.center);
		\draw [style=default] (26.center) to (27.center);
		\draw [style=default] (31.center) to (33.center);
		\draw [style=default] (32.center) to (34.center);
		\draw [style=default] (36.center) to (38.center);
		\draw [style=default] (37.center) to (39.center);
		\draw [style=default] (40.center) to (41.center);
		\draw [style=default] (42.center) to (43.center);
		\draw [style=default] (46.center) to (48.center);
		\draw [style=default] (47.center) to (49.center);
		\draw [style=default] (51.center) to (53.center);
		\draw [style=default] (52.center) to (54.center);
		\draw [style=default] (55.center) to (56.center);
		\draw [style=default] (57.center) to (58.center);
		\draw [style=default] (62.center) to (64.center);
		\draw [style=default] (63.center) to (65.center);
		\draw [style=default] (66.center) to (68.center);
		\draw [style=default] (67.center) to (69.center);
		\draw [style=default] (70.center) to (71.center);
		\draw [style=default] (72.center) to (73.center);
		\draw [style=default] (86.center) to (87.center);
		\draw [style=default] (88.center) to (89.center);
		\draw [style=default] (90.center) to (91.center);
		\draw [style=default] (92.center) to (93.center);
		\draw [style=default] (94.center) to (95.center);
		\draw [style=default] (96.center) to (97.center);
		\draw [style=default] (98.center) to (99.center);
		\draw [style=default] (100.center) to (101.center);
		\draw [style=default] (104.center) to (105.center);
		\draw [style=default] (108.center) to (109.center);
		\draw [style=default] (112.center) to (113.center);
		\draw [style=default] (116.center) to (117.center);
		\draw [style=default] (126.center) to (127.center);
		\draw [style=default] (133.center) to (135.center);
		\draw [style=default] (134.center) to (136.center);
		\draw [style=default] (138.center) to (140.center);
		\draw [style=default] (139.center) to (141.center);
		\draw [style=default] (142.center) to (143.center);
		\draw [style=default] (144.center) to (145.center);
		\draw [style=default] (149.center) to (151.center);
		\draw [style=default] (150.center) to (152.center);
		\draw [style=default] (153.center) to (155.center);
		\draw [style=default] (154.center) to (156.center);
		\draw [style=default] (157.center) to (158.center);
		\draw [style=default] (159.center) to (160.center);
		\draw [style=default] (161.center) to (162.center);
		\draw [style=default] (163.center) to (164.center);
		\draw [style=default] (165.center) to (166.center);
		\draw [style=default] (167.center) to (168.center);
		\draw [style=default] (171.center) to (172.center);
		\draw [style=default] (175.center) to (176.center);
		\draw [style=default] (179.center) to (180.center);
		\draw [style=default] (181.center) to (182.center);
		\draw [style=default] (183.center) to (184.center);
		\draw [style=default] (185.center) to (186.center);
		\draw [style=default] (187.center) to (188.center);
		\draw [style=default] (192.center) to (194.center);
		\draw [style=default] (193.center) to (195.center);
		\draw [style=default] (196.center) to (198.center);
		\draw [style=default] (197.center) to (199.center);
		\draw [style=default] (200.center) to (201.center);
		\draw [style=default] (202.center) to (203.center);
		\draw [style=default] (204.center) to (205.center);
		\draw [style=default] (206.center) to (207.center);
		\draw [style=default] (208.center) to (209.center);
		\draw [style=default] (210.center) to (211.center);
	\end{pgfonlayer}
\end{tikzpicture}

%% file: figures/decoupling.tikz
\begin{tikzpicture}
	\begin{pgfonlayer}{nodelayer}
		\node [style=none] (0) at (0, -1.5) {};
		\node [style=none] (1) at (-2.5, 1) {};
		\node [style=none] (2) at (2.5, 1) {};
		\node [style=none] (3) at (-2.5, 2) {};
		\node [style=none] (4) at (2.5, 5) {};
		\node [style=none] (5) at (0, -1) {};
		\node [style=none] (6) at (-2, 1) {};
		\node [style=none] (7) at (2, 1) {};
		\node [style=none] (8) at (-2, 2) {};
		\node [style=none] (9) at (2, 5) {};
		\node [style=none] (10) at (0, 0.5) {};
		\node [style=none] (11) at (-0.5, 1) {};
		\node [style=none] (12) at (0.5, 1) {};
		\node [style=none] (13) at (-0.5, 2) {};
		\node [style=none] (14) at (0.5, 5) {};
		\node [style=none] (15) at (0, -2) {};
		\node [style=none] (16) at (-3, 1) {};
		\node [style=none] (17) at (3, 1) {};
		\node [style=none] (18) at (-3, 2) {};
		\node [style=none] (19) at (3, 5) {};
		\node [style=none] (20) at (0, 0) {$\vdots$};
		\node [style=none] (21) at (-1.25, 1.5) {$\cdots$};
		\node [style=none] (22) at (1.25, 1.5) {$\cdots$};
		\node [style=none] (24) at (-3.5, 3.5) {};
		\node [style=none] (25) at (-3, 3.5) {};
		\node [style=none] (26) at (-0.5, 3.5) {};
		\node [style=none] (27) at (-4, 3.5) {};
		\node [style=none] (28) at (-3.5, 5) {};
		\node [style=none] (29) at (-3, 5) {};
		\node [style=none] (30) at (-0.5, 5) {};
		\node [style=none] (31) at (-4, 5) {};
		\node [style=none] (32) at (-1.25, 4.25) {$\cdots$};
		\node [style=none] (33) at (1.25, 4.25) {$\cdots$};
		\node [style=none] (34) at (-3.5, 2) {};
		\node [style=none] (35) at (-3.5, -0.5) {};
		\node [style=none] (38) at (-5, 2) {};
		\node [style=none] (39) at (-5, -0.5) {};
		\node [style=none] (40) at (-5.5, 2) {};
		\node [style=none] (41) at (-5.5, -0.5) {};
		\node [style=none] (42) at (-4.25, 0.75) {$\cdots$};
		\node [style={rect_l}, minimum width=3cm] (43) at (-3.25, 2.75) {$U_{MN}$};
		\node [style=none] (44) at (-6, 2) {};
		\node [style=none] (45) at (-6, -0.5) {};
		\node [style=none] (46) at (-4.5, 5) {};
		\node [style=none] (47) at (-4.5, 3.5) {};
		\node [style=none] (48) at (-5, 5) {};
		\node [style=none] (49) at (-5, 3.5) {};
		\node [style=none] (50) at (-5.5, 5) {};
		\node [style=none] (51) at (-5.5, 3.5) {};
		\node [style=none] (53) at (-6, 5) {};
		\node [style=none] (54) at (-6, 3.5) {};
		\node [style=none] (55) at (-2.5, 3.5) {};
		\node [style=none] (56) at (-2.5, 5) {};
		\node [style=none] (57) at (-2, 3.5) {};
		\node [style=none] (58) at (-2, 5) {};
		\node [style=none] (59) at (-3.25, -0.75) {};
		\node [style=none] (60) at (-6.25, -0.75) {};
		\node [style=none] (61) at (-4.75, -1.5) {$\ket{\psi_{\text{anc}}}_M$};
		\node [style=none] (62) at (0.25, 5.25) {};
		\node [style=none] (63) at (3.25, 5.25) {};
		\node [style=none] (64) at (1.75, 6) {$R$};
		\node [style=none] (65) at (-4.75, 5.25) {};
		\node [style=none] (66) at (-2.75, 5.25) {};
		\node [style=none] (67) at (-3.75, 6) {$A$};
		\node [style=none] (68) at (2.25, -1.5) {$\ket{\Phi^+}_{NR}$};
	\end{pgfonlayer}
	\begin{pgfonlayer}{edgelayer}
		\draw [style=default] (1.center) to (0.center);
		\draw [style=default] (0.center) to (2.center);
		\draw [style=default] (1.center) to (3.center);
		\draw [style=default] (2.center) to (4.center);
		\draw [style=default] (6.center) to (5.center);
		\draw [style=default] (5.center) to (7.center);
		\draw [style=default] (6.center) to (8.center);
		\draw [style=default] (7.center) to (9.center);
		\draw [style=default] (11.center) to (10.center);
		\draw [style=default] (10.center) to (12.center);
		\draw [style=default] (11.center) to (13.center);
		\draw [style=default] (12.center) to (14.center);
		\draw [style=default] (16.center) to (15.center);
		\draw [style=default] (15.center) to (17.center);
		\draw [style=default] (16.center) to (18.center);
		\draw [style=default] (17.center) to (19.center);
		\draw [style=default] (31.center) to (27.center);
		\draw [style=default] (28.center) to (24.center);
		\draw [style=default] (29.center) to (25.center);
		\draw [style=default] (30.center) to (26.center);
		\draw [style=default] (35.center) to (34.center);
		\draw [style=default] (39.center) to (38.center);
		\draw [style=default] (41.center) to (40.center);
		\draw [style=default] (45.center) to (44.center);
		\draw [style=default] (47.center) to (46.center);
		\draw [style=default] (49.center) to (48.center);
		\draw [style=default] (51.center) to (50.center);
		\draw [style=default] (54.center) to (53.center);
		\draw [style=default] (56.center) to (55.center);
		\draw [style=default] (58.center) to (57.center);
		\draw [style=brace] (59.center) to (60.center);
		\draw [style=brace] (62.center) to (63.center);
		\draw [style=brace] (65.center) to (66.center);
	\end{pgfonlayer}
\end{tikzpicture}

%% file: sections/numerical.tex
\section{A Numerical Study}
\label{sec:numerical}

We now present a (non-exhaustive) numerical analysis of the NoRA tensor network using the Clifford stabilizer formalism discussed previously. Our primary focus is the scaling of the (relative) code distance with $N$, and how it differs between having the space of ground states scale with $L$ and having it fixed.

The stabilizer simulation used to generate the following data was written in Python 3.10.4 using Numpy 1.21.6 (linear algebra) \cite{harris_array_programming_numpy_2020} and Galois 0.1.1 (finite field arithmetic) \cite{hostetter_galois_performant_numpy_2020}, and is based on the projective symplectic representation discussed in appendix \ref{sec:phase_space}. The algorithm used to randomly sample symplectic matrices for Clifford operators is based on \cite{koenig_how_efficiently_select_2014}, but was generalized to work for any choice of qudit dimension $d$ that is a power of an odd prime. The complete code can be found at
\url{https://github.com/vbettaque/qstab}.

The datasets used in this paper were generated using a 2021 MacBook Pro with M1 Pro processor and 16 GB RAM, and can be shared upon request. If the computation involved random sampling, an average of 1000 samples is displayed together with the error on the mean\footnote{Note that occasionally the error on the mean is so small that it is not visible in the figures.}. In general we also chose a qudit dimension of $d=3$, a growth rate of $r=2$ per layer, and a (naive) layer circuit growth rate of $q=2$. 

\subsection{Fixed Ground Space Size}
\label{sec:gs_fixed}

We begin our analysis with the case where the size of the ground space is fixed. The other case, where the size of the ground space grows with $L$, more closely resembles SYK, but the fixed size case is also interesting as a starting point and for the codes it produces. In such cases, the rate $k/N$ of the code approaches zero exponentially fast with the total layer number. However, the complexity still increases exponentially in $L$ according to
\begin{equation}
    \text{total gates} = \frac{D}{q} \frac{r}{r-1} \cdot N + \mathcal{O}(\log N),
\end{equation}
suggesting that distances scaling with $N$ should be achievable. The vanishing rate is also not inherently problematic as this is also true for other error-correcting codes like the $[[A^2, 2, A]]$ toric code \cite{kitaev_fault_tolerant_quantum_2003}.

\subsubsection{Entanglement Entropy}
\label{sec:entanglement_entropy}

\begin{figure}[htb]
    \centering
    \includegraphics{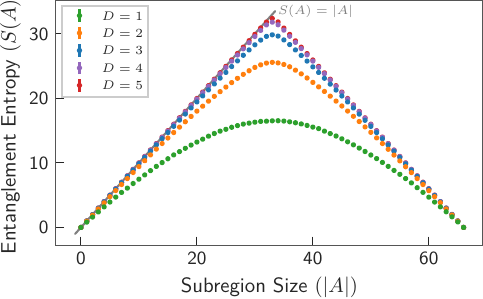}
    \caption{Average entanglement entropy $S(A)$ with regard to the subregion size $|A|$, corresponding to a NoRA-prepared stabilizer state ($k = 2, L = 6$) with different layer circuit depths $D$. Volume-law scaling of $S(A)$ for small $|A|$ is already possible when $D = 2$, but bigger subregions (up to a size of $N/2 = 33$) also exhibit a volume scaling when $D$ is larger. Once $|A|$ is larger than $N/2$, the entanglement entropy decreases again for any choice of $D$, forming a symmetric Page curve.}
    \label{fig:volume_law}
\end{figure}

The first part of our analysis deals with directly confirming the expected volume-law entanglement of the average NoRA-prepared stabilizer state. For any state to have that property, the (von-Neumann) entanglement entropy $S(A)$ associated to a random subregion $A$ of the state should scale with the size $|A|$ of the region, at least for the right parameters and as long as one has $|A| < N/2$. Figure \ref{fig:volume_law} depicts (approximately) that behavior for $k=2$, $L=6$ and $D > 1$. The case of $D = 1$ is also shown, but the corresponding average entropy fails to scale linearly with $|A|$ at all.

Figure \ref{fig:volume_law} also shows that once the size of $A$ increases beyond $N/2$, the entropy shrinks again. This is due to the complementary subregion $\overline{A}$ now being smaller than $A$, and both regions sharing the same entanglement spectrum. Overall the entanglement entropy $S(A)$ therefore follows a symmetric \emph{Page curve}, as expected.

\subsubsection{Code Distance}
\label{sec:fixed_distance}

\begin{figure}[htb]
    \centering
    \includegraphics{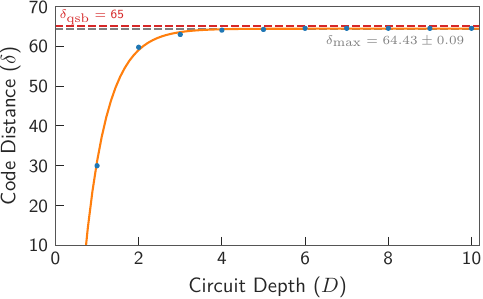}
    \caption{The average code distance $\delta$ with regard to the circuit depth $D$ of the NoRA network with a fixed ground space ($k=2, L=7$). Due to the approximately exponential trend of the data, the average distance is already close to its saturated maximum of $\delta_{\max} = 64.43 \pm 0.09$ if the layer circuits have a depth of $D=3$. Larger depths therefore provide little to no improvement. However, the maximum average distance is still several standard deviations less than the theoretical maximum provided by the quantum singleton bound $\delta_{\textrm{qsb}}$, though we expect them to be somewhat closer (but not necessarily equal) at large $D$ and $L$.}
    \label{fig:d_against_D}
\end{figure}

We now turn to determining how the average code distance depends on the layer-circuit depth $D$. This is of interest to us since for error correction we want to choose $D$ to be as small as possible to reduce the circuit complexity, while still having $\delta$ as large as possible (i.e.\ \emph{saturated}) on average. Looking at figure \ref{fig:d_against_D}, that seems to be the case for $D_{\text{sat}} = 2,3,4$, depending on the tolerated margin of error between $\delta$ and $\delta_{\max}$. The case of $D_\text{sat} = 1$ not coming close to maximizing the distance coincides with the lack of volume-law entanglement in figure \ref{fig:volume_law} for the same depth, albeit for different choices of $L$.

In general one would assume that $D_{\text{sat}}$ depends on $L$ as well as all other parameters. However, we can argue that $D_{\text{sat}}$ is largely independent of $L$ and should only strongly depend on $d$, $q$ and $r$. As is shown in appendix \ref{sec:growth_factor}, the weight of an operator string increases on average by a factor of $g^{D_{\text{sat}}}$ (where $g \approx q \cdot (d^2 - 1)/d^2$ only depends strongly on $d$ and $q$) when subjected to a single $q$-local Clifford with depth $D_{\text{sat}}$. And since the distance of a typical stabilizer code is expected to scale like its operator weight, we can adopt the same arguments here\footnote{The same reasoning is used to estimate the distance scaling in section \ref{sec:syk_distance_weight}.}. This means that $D_{\text{sat}} \approx \log_g (n_{\ell} / w_{\ell-1})$ should estimate the minimum depth needed for both weight and distance saturation.

In the case of NoRA with $k$ fixed we have $n_{\ell}/n_{\ell-1} = (k + r^{\ell})/(k + r^{\ell - 1}) \approx r$ for $r^{\ell} \gg k$, meaning the system increases by a roughly constant factor at each layer. Given that we start in a steady state, i.e.\ with a Pauli string with close to maximum weight ($w_{\ell -1} \approx n_{\ell - 1})$, then for the string in the subsequent layer to also have maximum weight we require that $w_{\ell} \approx n_{\ell} \approx g^{D_\text{sat}} \cdot n_{\ell - 1}$ and hence
\begin{equation}
\label{eq:D_sat_NoRA}
    D_\text{sat} \approx \log_g(r), 
\end{equation}
which does not depend strongly on $L$ (or $\ell$). For $d = 3$ and $q = r = 2$ this gives $D_\text{sat} \approx 1.2$ which rounds up to 2, the observed lower bound for the occurrence of volume-law scaling. Additional numerical evidence for this heuristic with regard to different choices of $L$ is also provided by figure \ref{fig:syk_weight_diff} in section \ref{sec:weight_results}. Overall we therefore can have $D_\text{sat}$ fixed for different sizes of the tensor network without expecting a significant impact on the relative code distance $\delta / N$.

\begin{figure}[htb]
    \centering
    \includegraphics{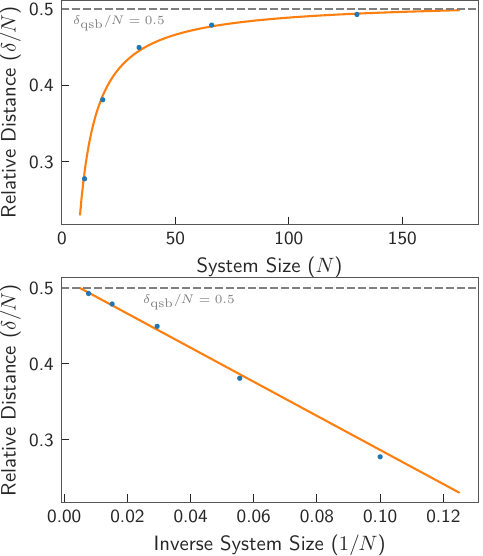}
    \caption{Average relative code distance $\delta/N$ with regard to the system size $N = 2 + 2^L$ and its inverse $1/N$, with $D = 4$. Even though the depicted data trends are only approximate, they do suggest approximate convergence of the relative distance towards the (relative) quantum singleton bound $\delta_{\textrm{qsb}}/N = 0.5$ for larger $N$.}
    \label{fig:dn_n_inv}
\end{figure}

Figure \ref{fig:d_against_D} also shows that the tensor network can (on average) achieve distances that are quite close to the theoretical maximum:
\begin{equation}
    \delta_{\text{qsb}} = \frac{N - k}{2} + 1 = \frac{N}{2} = 2^{L-1} + 1.
\end{equation}
This maximum is assumed if the quantum singleton bound $N - k \geq 2 (\delta-1)$ becomes an equality. Reaching this \emph{saturation limit} (or at least coming close to it) requires states with volume law entanglement, thus verifying our previous expectations. It would be interesting to understand how close the average code distance comes to $\delta_{\text{qsb}}$ as a function of $L$ and $D$. However the computing time scales exponentially with $N$ and therefore double-exponentially with $L$, making it more difficult to gather data for larger system sizes. But for now our results do indeed suggest a possible approximate distance saturation with more layers, as shown in figure \ref{fig:dn_n_inv} for one specific example\footnote{Note that the relative singleton distance $\delta_{\text{qsb}}/N$ is not necessarily independent of $N$. This is only the case here because we chose $k=2$. An example for a $N$-dependent relative distance is given in the next section.}. We say approximate because \eqref{eq:D_sat_NoRA} implies that for reasonable choices of $D$ we expect the system to reach a steady state after a certain number of layers, meaning that for subsequent layers the scrambling rate of the finite-depth circuit and the rate of new thermal qudits form an equilibrium and thus keep the relative distance constant. Depending on the choice of parameters, this equilibrium does not necessarily have to coincide with $\delta_{\text{qsb}}$. However, we expect this to be the case for unreasonably large scrambling rates $g^D \gg r$ due to the network dynamics being dominated by the upper-most finite-depth circuit $D_L$.

\begin{figure}[htb]
    \centering
    \includegraphics[width=\linewidth]{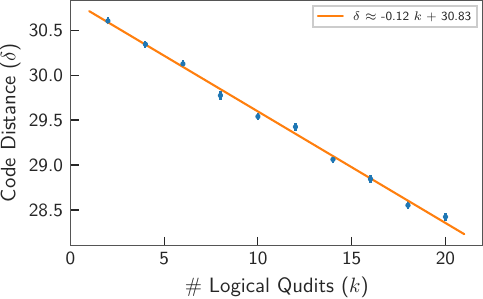}
    \caption{Change of the code distance with regard to increasing ground space size $k$, and keeping the number of layers fixed to be $L = 6$ with circuit depth $D = 3$. The approximately linear decrease of the distance is indicative of the tensor network being able to create volume-law entanglement. }
    \label{fig:d_against_k}
\end{figure}

Finally we consider how the code distance $\delta$ scales when the number of logical qudits $k$ is increased while keeping the number of layers $L$ fixed. Doing so provides another heuristic as to whether the tensor network exhibits volume-law entanglement or not, since we expect a linear decrease of the entanglement entropy and therefore distance with increasing $k$ in that case. As shown in figure \ref{fig:d_against_k} this seems to be indeed the case on average and for our choice of parameters.

\subsubsection{Stabilizer Weights}
\label{sec:weight_results}

Besides the average code distance of our tensor network ansatz, it is also interesting to consider the weight distribution of the (naive) stabilizer basis describing the code. For the purpose of performing error correction, having a low-weight code (meaning a code with a low weight generating set for the stabilizer group) is desirable since it means the syndrome can be obtained by measuring low-weight operators. If many of the stabilizers are high weight, it might be infeasible to measure the entire syndrome before too many errors accumulate. Moreover, the commuting projector Hamiltonian whose ground space coincides with the code space is only local (few-body) if the code is low-weight.

\begin{figure}[htb]
    \centering
    \includegraphics{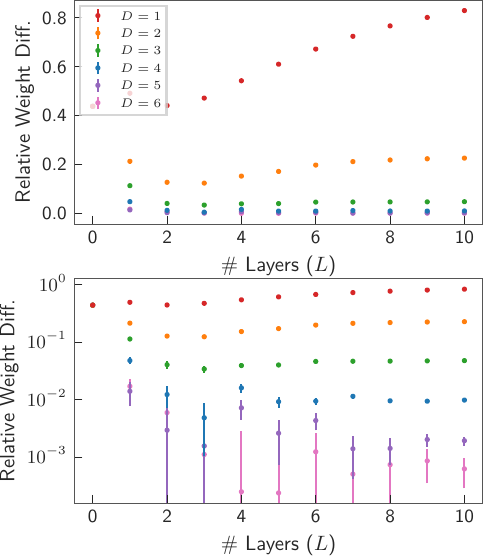}
    \caption{(Logarithmic) relative difference between the actual weights of a single Pauli string and their expected maximum of $N \cdot (d^2 - 1)/d^2$ for the tensor network ansatz with a variable number of layers $L$ and depths $D$ ($k=2$). Already for a circuit depth of $D=2$ does the relative difference seemingly converge towards a constant value, with larger depths resulting in even faster convergence and smaller differences. This shows that the relative weight (and therefore the ideal circuit depth to maximize the code distance) does only depend strongly on $d$, $q$ and $r$, not $L$.}
    \label{fig:syk_weight_diff}
\end{figure}

To analyze the whole stabilizer basis, we first consider how the tensor network affects the weight of a single Pauli string with unit weight. The non-identity operator is here at the beginning of the string, which means that it is acted on non-trivially by all layers of the circuit. The resulting averaged weight evolution is depicted for different choices of $D$ in figure \ref{fig:syk_weight_diff} in terms of its relative difference to the expected maximum weight which is given at each layer $\ell$ by
\begin{equation}
    w_{\ell}^{\text{sat}} = \frac{d^2 - 1}{d^2} \cdot n_{\ell}.
\end{equation}
What can be seen is that for all choices of $D>1$ the weight differences reach an equilibrium\footnote{We expect this to happen for $D=1$ as well, however at larger $L$ and with a large relative difference compared to the other circuit depths. More on that later in this section.}, barely changing for later layers. The same circuit depth therefore always approximately produces the same relative weight, regardless of the actual number of layers $L$. We already used this argument in section \ref{sec:fixed_distance} to argue that the minimal depth $D$ to achieve a good distance $\delta$ does not depend on $L$ because distance and weight usually have correlating behaviors.

\begin{figure}[htb]
    \centering
    \includegraphics{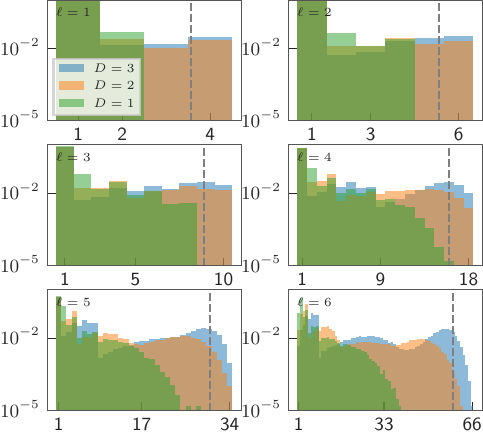}
    \caption{Relative stabilizer weight distributions at each layer of the tensor network ansatz with $L=6$ total layers. The green distributions correspond to circuit depth $D=1$, the orange ones to $D=2$ and the blue ones to $D=3$. The dashed lines indicate where the expected maximum weight averages $w_{\ell}^{\max}$ are located for each layer. Note that as expected the distribution corresponding to $D=1$ does not converge towards saturation due to the rate $r$ of new qudits being added outweighing the scrambling rate $g^D$ of the circuits at each layer.}
    \label{fig:stab_weights}
\end{figure}

However, when considering a complete set of stabilizer states, not all of them are acted on non-trivially immediately since they might correspond to thermal qudits introduced only in later layers. Take for example the number of thermal qudits introduced in the final layer of a NoRA circuit, which experience the least scrambling of all degrees of freedom but make up  $r^{L-1}$ of the total $k + r^L$ qudits. For $k = r = 2$ this is close to 50\%. Unlike an arbitrary random circuit with equivalent total depth, where all stabilizer basis elements experience the same amount of scrambling and are therefore expected to have equally high weights, we predict the stabilizer weights of a NoRA circuit to obey a multimodal distribution with dominant peaks around both 1 and $w_{L}^{\text{sat}}$. This is indeed approximately the case as seen from the specific example depicted in figure \ref{fig:stab_weights}. It also shows that with increased circuit depth $D$ more and more basis weights approach saturation, as expected.

Note though that for $D=1$ all stabilizers fail to come anywhere close to maximum weight. This aligns with the predictions that are coming up in section \ref{sec:syk_distance_weight}, where we suggest that some sort of phase transition should occur in the relative stabilizer weight distribution (and hence distance) when going from the regime of $g^D < r$ to $g^D > r$, and considering large $L$. In the former case we expect the average stabilizer weight to be small to negligible compared to the total size, while in the latter case we predict complete weight saturation for all elements. For the specific example in figure \ref{fig:stab_weights} we assumed $g = q \cdot (d^2 - 1)/d^2 = 16/9 < 2$ (as shown in appendix \ref{sec:growth_factor}) and $r = 2$, meaning that we should have $g^D < r$ for $D=1$ and $g^D > r$ for $D>1$. And since the relative weights for $D=1$ are comparatively small, this indicates that this transition does indeed take place. In the future we intend to explore this behavior in more detail by looking at other examples in the parameter space.

Comparing the weight analysis with our results from the previous section we can therefore conclude that the stabilizer bases with the lowest weights and highest distances are achieved when choosing $D=3$ as the layer circuit depth. Choosing $D=1$ could also be beneficial though at a significant cost of distance. Either way, the relative number of high-weight stabilizers is significant, thought we expect there to be potential for further reducing the weights, as will be explained in section \ref{sec:weight_reduction}.

\subsection{Enabling Ground Space Scaling}

To model a situation like SYK where the number of ground state qudits is proportional to the total number of qudits, and where we have a thermodynamic limit where both numbers go to infinity, we want to take $L \sim \log_r k$, although this is not an integer in general. So consider as a simple model the case where $k=r^a$ and $L=a+b$ for two integers $a$ and $b$. Then the total number of qudits is 
\begin{equation}
    N = r^a + r^{a+b},
\end{equation}
and the ratio between ground state qudits and the total number of qudits is therefore
\begin{equation}
    \frac{k}{N} = \frac{1}{1 + r^b},
\end{equation}
which is independent of $a$. The limit $a \rightarrow \infty$ can thus be viewed as a thermodynamic limit in which $N$ and $k$ diverge but with a fixed finite ratio. By varying $b$ we can then adjust the relative number of ground state qudits. 

Many of our previous arguments that assume the number of layers to be fixed therefore apply here as well and will not be repeated. Of primary interest to us is therefore how our code's distance and weight scale with $N = r^a + r^{a+b}$ where $a$ increases and $b$ is fixed. In addition to the previously made choices for $d$, $q$ and $r$, we also assume $D=3$ and $b = 1$ for the following examples.

\subsubsection{Code Distance}

\begin{figure}[htb]
    \centering
    \includegraphics[width=\linewidth]{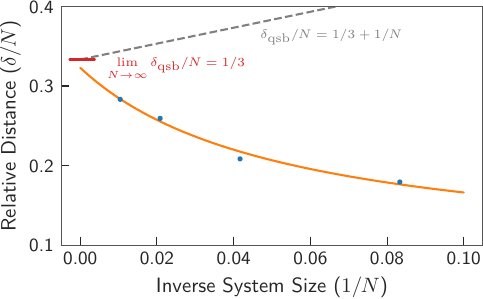}
    \caption{Average relative code distance $\delta/N$ with regard to the inverse system size $1/N$ for the tensor network ansatz with ground space scaling. Both the quantum singleton bound (in grey) and the approximate trend of the generated data (in orange) could coincide in the limit of $N \rightarrow \infty$ ($a \rightarrow \infty$), where we have $\delta / N = 1/3$. However more data is needed to be able to prove this. Overall, better relative distances can only be achieved by increasing $b$ at the cost of reducing the rate. }
    \label{fig:d_rel_scaling}
\end{figure}

Before considering explicit simulations, we can again use the quantum singleton bound to find an upper bound for the expected relative code distances. This bound turns out to be
\begin{align}
\begin{split}
    \frac{\delta_{\text{qsb}}}{N} &= \frac{(N-k)/2 + 1}{N} \\
    &= \frac{1}{2 \, (1 + r^{-b})} + \frac{1}{N} \\
    &= \frac{1}{3} + \frac{1}{N} \quad (b = 1, r = 2),
\end{split}
\end{align}
which unlike the fixed case is necessarily dependent on $N$, although only weakly at large $N$. For large system sizes the relative distance therefore approaches the fixed value of $1/3$ for $b=1$ (and $r=2$). Comparing this to numerical approximations of the average distance and its trend as shown (in orange) in figure \ref{fig:d_rel_scaling}, we can see that both trends might coincide in that very limit, or at least come close.  

\subsubsection{Stabilizer Weights}

\begin{figure}[htb]
    \centering
    \includegraphics{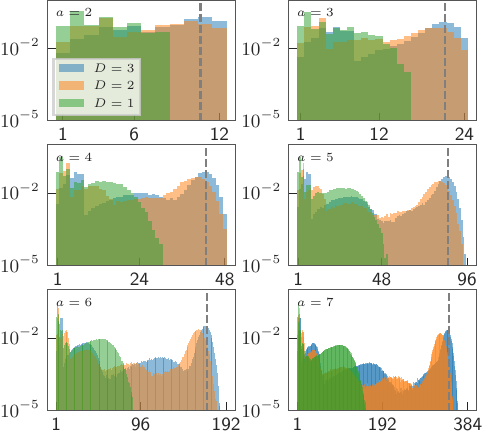}
    \caption{Relative stabilizer weight distributions of the tensor network ansatz with scaling ground space and different choices of $a$. The green distributions correspond to circuit depth $D=1$, the orange ones to $D=2$ and the blue ones to $D=3$. These distribution are not dissimilar to what we already encountered in \ref{fig:stab_weights} for the individual layers of a single circuit, highlighting the self-similarity of our ansatz.}
    \label{fig:syk_stab_weights}
\end{figure}

As seen in figure \ref{fig:syk_stab_weights} the weight distributions for the SYK-like NoRA model don't differ significantly from the case of a fixed ground space. The only significant difference lies in the origin of the distributions: In the case of a scaling ground space we extracted the weights from circuits with different choices for $a$, while in the fixed case we depicted the weights at each layer of a single circuit. That both cases nevertheless produce similar figures is due to the fact that our tensor network ansatz exhibits self-similarity.

It remains to be shown that this trend occurs for different choices of $g$ and $r$ and continues as expected for larger $a$ and $b$. We are also interested in exploring the phase transition at hand in the limit of large $L$.
Those are things we intend to explore in future work.

\subsection{Summary}

Through extensive numerical simulations, we verified that the stabilizer codes obtained from the random Clifford layers indeed can have volume-law entanglement, and relative distance approaching a non-zero constant in the thermodynamic limit $N\rightarrow \infty$. This corresponds to distance proportional to $N$, which in turn is also a heuristic for the presence of volume-law entanglement. The relative distance depends on the model parameters, especially the depth $D$, with the result coming close to the relative singleton bound in both the fixed $k$ case and the $k \propto N$ case as $D$ is increased. We also found a broad distribution of stabilizer weights, with a few high weight stabilizers coming from the near-IR thermal qudits and a larger number of low weight stabilizes coming from the near-UV thermal qudits. Our architecture with random layers is therefor capable of producing a family of codes indexed by $N$ with non-vanishing relative distance and rate at the cost of having some high weight stabilizers (although significantly fewer than in a fully random code).

%% file: sections/code_struct.tex
\section{Analysis of the SYK-Inspired Code}
\label{sec:code_structure}

We now consider in more detail the properties of the SYK inspired code with $k=r^a$ and $L=a+b$ for two integers $a$ and $b$. Recall that the total number of qudits is 
\begin{equation}
    N = r^a + r^{a+b},
\end{equation}
and the ratio between ground state qudits and the total number of qudits (i.e. the rate) is therefore
\begin{equation}
    \frac{k}{N} = \frac{1}{1 + r^b},
\end{equation}
which is independent of $a$. The thermodynamic limit $a \rightarrow \infty$ gives a family of codes with non-zero rate. We already established in Section~\ref{sec:numerical} that this code can be highly entangled. It is also interesting to consider its complexity.

In this case, the complexity sum \eqref{eq:circuit_complexity} can be rewritten as
\begin{equation}
    \text{total gates} = \frac{D}{q \cdot (1 + r^b)} \cdot N \log_r N + \mathcal{O}(N).
\end{equation}
This leading $N \log N$ scaling with the total number of degrees of freedom can be compared to holographic complexity conjectures applied to JT gravity~\cite{brown_case_missing_gates_2019}; one also gets $N \log N$ by studying, for example, the volume (length) of the wormhole dual to the thermofield double state with temperature of order $1/N$. The key point is that the throat of the wormhole is long, of order $\log N$, at this temperature. Hence, the circuit complexity of our SYK-inspired encoding also resembles that obtain from holographic models dual to SYK.

For the estimates discussed below, we continue to assume that the layers are composed of random 2-qudit Clifford gates applied to random pairs of qudits. We caution that this is certainly not correct for the actual SYK model: the gates must act on fermionic degrees of freedom and will not be Clifford (or the fermionic analogue of Clifford) generically. Here we continue to focus on the Clifford case for ease of analysis and for its interpretation in terms of an exact quantum error correcting code. Below we comment briefly on the potential similarities and differences with the actual SYK model.

\subsection{Distance Estimate and Stabilizer Weights}
\label{sec:syk_distance_weight}

We know the rate of our SYK-inspired code. To estimate the distance, we need to understand how logical operators grow as they pass from the IR to the UV. Let us assume that a typical operator grows in size by a factor of $g^D$ after passing through one layer (i.e. being conjugated by that layer unitary), up to a maximum size set by the total number of qudits. A way to estimate $g$ when the layer unitary is a random Clifford circuit can be found in appendix \ref{sec:growth_factor}. At the same time, the number of qudits is also growing, going from $k+r^{\ell-1}$ to $k+r^\ell$. The distance depends on whether the size of operators grows faster or slower than the number of qudits. Note that we saw already a manifestation of this competition in the discussion in Section~\ref{sec:numerical}; here we explain in more detail the issues.

From a given random circuit layer, we expect operators to grow by a factor of $g^D$ provided they are not close to maximum weight. If they are close to maximum weight, then they will grow by a reduced factor. We must compare this operator growth to the rate of qudit increase. The ratio $R_\ell$ between the number of qudits in successive layers is 
\begin{equation}
      R_\ell =  \frac{k+r^\ell}{k+r^{\ell-1}} = 1 + \frac{r-1}{1 + k/r^{\ell-1}}, 
\end{equation}
which monotonically increases with $\ell$. As logical operators evolve from layer to layer into the UV, the relative weight of the operator either increases or decreases depending on whether $g^D > R_\ell$ or $g^D < R_\ell$. The dynamics of this process, iterated over all $L$ layers, gives an estimate for the size of non-trivial logical operators.

\subsubsection{Warmup: Small Fixed \textit{k}}

To illustrate the key competition, consider first the case in which $k$ is small and fixed. In this case, the ratio $R_\ell \rightarrow r$ as $\ell$ increases, so most of the evolution corresponds to a fixed ratio of $r$. In terms of the parameters above, we can achieve this regime by taking $b$ large at fixed $a$.

Suppose $g^D > r$. Then operator growth is the fastest process and logical operators will reach saturation. In this case, we expect the distance to be linear in $N$. It will not exactly saturate the singleton bound, but it may come close for large $D$.

Now suppose $g^D < r$. In this case, we are adding qudits faster than operators can grow, so the logical operators are ultimately supported on a dilute fraction of all the sites. Indeed, the size of a typical logical operator will be $g^{DL}$, whereas the total number of qudits is $N =  r^L (1 + r^{-b}) \approx r^L$. Expressed in terms of $N$, the size of a typical logical operator is
\begin{equation}
    g^{DL} \sim N^c
\end{equation}
where $c = \frac{\ln g^D}{\ln r} < 1$. Hence, we expect a distance that scales as a sublinear function of $N$.

\subsubsection{SYK-Like Scaling}

Now we turn to the case where $k = r^a$ is large and $b$ is fixed. Here, when $\ell$ is small, the $R_\ell$ ratio is close to one and the number of qudits is barely increasing from layer to layer. In this regime, operator growth is completely dominant. In contrast, at the most UV layer, where $\ell=L=a+b$, the ratio is
\begin{equation}
   R_L = 1 + \frac{r-1}{1+r^{1-b}} < r. 
\end{equation}

Suppose $g^D > R_L$. Then operator growth always dominates over qudit growth. However, because the initial number of qudits (the ground state qudits) is large, we still have to compare the total operator size, $g^{D L}$, to the total number of qudits, $N = r^L ( 1+ r^{-b})$. We see again that if $g^D >r$, then this naive estimate gives an operator weight larger than $N$, meaning that the operator growth actually saturated at something proportional to $N$. If $g^D < r$, then we are again in the situation where $g^{DL} \sim N^c$.

Suppose $g^D < R_L$. Then there will be some layer $\ell^*$ such that operator growth and qudit growth switch dominance as $\ell$ increases through $\ell^*$. We may approximately determine this crossover scale from
\begin{equation}
    g^D = R_{\ell^*},
\end{equation}
noting that this $\ell^*$ is not typically an integer. In the thermodynamic limit $a \rightarrow \infty$, we must have $\ell^* = a + b^*$ for some constant $b^*$ since the ratio $R_\ell$ is essentially unity until $r^{\ell}$ is comparable to $k$.

Now between $\ell=1$ and $\ell = \ell^*$, logical operators will grow faster than the number of qudits. Assuming they don't reach saturation, they will grow by roughly a factor of $g^{D\ell^*}$. By contrast, the number of qudits at layer $\ell^*$ is
\begin{equation}
    n_{\ell^*} = r^a (1 + r^{b^*}),
\end{equation}
so the ratio of operator size to number of qudits is
\begin{equation}
    \left( \frac{g^D}{r} \right)^a \frac{g^{D b^*}}{1+r^{b^*}}.
\end{equation}
This ratio vanishes as $a\rightarrow \infty$ since we are assuming that $g^D < R_L$ and $R_L < r$. Hence, $g^{D\ell^*} \sim (n_{\ell^*})^c$ as above.

There are a fixed number of layers from $\ell^*$ to $L$ since $b$ and $b^*$ are fixed as $a\rightarrow \infty$. Therefore operators and the number of qudits grow by an additional factor independent of $N$ from $\ell^*$ to $L$. Hence, the scaling of $g^{DL}$ with $N$ is the same as the scaling of $g^{D\ell^*}$ with $n_{\ell^*}$, that is $g^{DL} \sim N^c$.

\subsubsection{Stabilizer Weights}

We expect that the stabilizer weights will display a similar pattern as in Figure~\ref{fig:stab_weights}. In particular, a non-zero fraction of all the stabilizers will have constant weight. These arise from the UV most layer. Then as we descend in the network towards the IR, there are fewer stabilizers but of increasing weight. In particular, there are at least a few stabilizers of very high weight, similar to the weight of logical operators. 

\subsection{Comparison to SYK} 

We now compare features of the SYK-inspired code to those of the actual SYK model. To be precise, we will compare a particular realization of the SYK Hamiltonian (with $q=4$), $H_{\text{SYK}}$, with a particular realization of the toy code Hamiltonian, $H_{\text{code}}$, for the SYK-inspired code (see Section~\ref{sec:arch}). (It is also interesting to consider supersymmetric generalizations~\cite{fu_publisher_note_supersymmetric_2017}.)
\begin{itemize}

   \item{\textbf{Hamiltonian structure:}} $H_{\text{SYK}}$ is composed of $O(N^4)$ weight-$4$ fermion terms (all possible such terms). These terms do not all commute and they enter $H_{\text{SYK}}$ with random coefficients. $H_{\text{code}}$ is composed of $O(N)$ commuting terms with fixed coefficients. The weight of the terms varies, with many having low-weight but a significant fraction having high weight, comparable to the distance of the corresponding code. The two ensembles are therefore by definition not directly analogous, but are expected to have analogous properties.
   
    \item{\textbf{Ground space:}}  $H_{\text{SYK}}$ has $e^{s_0 N}$ approximate ground states which are approximately degenerate with level spacing $e^{-\alpha N}$. $\alpha$ and $s_0$ are constants, independent of $N$. Similarly, $H_{\text{code}}$ has $d^k = e^{ \frac{\ln d}{1+r^b} N}$ exactly degenerate ground states.
    
    \item{\textbf{Low-temperature thermodynamics:}} The SYK model has a low temperature heat capacity proportional to temperature $T$. Similarly, the parameters of $H_{\text{code}}$ can be chosen so that its low temperature heat capacity is proportional to $T$.
    
    \item{\textbf{Fine-tained spectrum:}} The fine-grained energy spectrum of $H_{\text{SYK}}$ is random-matrix-like~\cite{cotler_black_holes_random_2017, saad_semiclassical_ramp_syk_2019}. The fine-grained energy spectrum of $H_{\text{code}}$ is not random-matrix-like because $H_{\text{code}}$ is a commuting projector Hamiltonian.
    
    \item{\textbf{Entanglement:}} Both models feature Hamiltonian eigenstates with volume-law entanglement. The entanglement spectrum will, however, be quite different between the two kinds of states. In particular, eigenstates of $H_{\text{code}}$, being stabilizer states, have a flat entanglement spectrum.
    
    \item{\textbf{Complexity:}} We only have estimates here. Using the duality to JT gravity and holographic complexity/geometry conjectures, the circuit complexity of the SYK approximate ground states is estimated to be $O(N \ln N)$. We have an explicit estimate (and upper bound) of $O(N \ln N)$ for circuit complexity of the ground space of $H_{\text{code}}$.

\end{itemize}

The many similarities between $H_{\text{SYK}}$ and $H_{\text{code}}$ are the basis for our conjecture that the architecture in Figure~\ref{fig:arch} has the potential to describe the physics of the SYK model once the tensors in the network have been adapted to a particular SYK instance, for example, using a variational approach. However, there are also crucial differences between the two. Two that stand out are the different scalings of the weights of Hamiltonian terms with system size and the exact versus approximate nature of the ground state degeneracy. The fine-grained energy spectrum is also very different in the two cases. Thus, it will be informative in the future to explore our network architecture as a variational ansatz for the SYK ground space.

\subsection{SYK Ground Space as an Approximate Code}

Here we want to comment on another possibility raised by the similarities above. For $H_{\text{code}}$, we have seen explicitly that the ground space can be viewed as an error correcting code with constant relative distance and constant rate (provided $D$ is big enough). In particular, it is an exact stabilizer code. This naturally raises the possibility that the approximate ground space of the SYK model could have interesting properties as an approximate quantum error correction code.\footnote{The network architecture presented here was first considered by one of the authors in fall 2019 during their stay at the Institute for Advanced Study and later presented in preliminary form, along with the potential code interpretation, in January 2020 at UCSB. Independently, the code properties of SYK in the thermal regime have been studied~\cite{chandrasekaran_quantum_error_correction_2022}.}

Thus we consider a code defined by the full approximate ground space of some particular $H_{\text{SYK}}$ realization. By construction this code has a constant rate as $N\rightarrow \infty$ which is given by ground state entropy density $s_0$. This code is not a stabilizer code, but it does have a sort of ``low weight'' definition via the SYK Hamiltonian. 

What is not immediately clear is the distance of this code. Moreover, since the code is approximate, we must specify precisely what we mean by the distance. We will defer a full discussion to a future work, but here let us note that if the architecture in Figure~\ref{fig:arch} does indeed provide a good approximation to the ground space of the SYK realization, then the same kind of scaling analysis discussed above for the random Clifford code would also provide an estimate for the operator size of logical operators.

In this case, it would be important to understand the analog of $r$ and $g^D$ in the SYK case. As one approach, we could fix $r=2$ and then adjust the layer circuits so that we get a good approximation to the ground space. The parameter $g^D$ would then be determined by the properties of these circuit. A simple random operator growth model may be too crude to capture the detailed physics, but continuing with this estimate for now, if the resulting $g^D$ were greater than $r$, then we have logical operators of weight proportional to $N$ and potentially distance proportional to $N$. Alternatively, if $g^D < r$, then the distance could be some power of $N$, $N^c$. It would be interesting to understand which of two cases is realized; this should be related to the spectrum of the scaling dimensions in the theory since these are related to the mixing properties of the scaling superoperator~\cite{vidal_class_quantum_many_2008}. Given the relatively low scaling dimension of the fermion operators, it may be that one is effectively in the $g^D<r$ regime.

\subsection{Summary}

We gave analytical estimates of the distance for a family of SYK-inspired codes in the thermodynamic limit of many qudits. This code family shares a number of similarities with known properties of the actual SYK model, although there are crucial differences as well. Viewing the approximate ground space of SYK as an approximate quantum code, the analysis of the SYK-inspired model suggests that the actual SYK ground space code, which has constant rate as $N\rightarrow \infty$, could have a distance $N^c$ for some constant $0<c\leq 1$.

%% file: sections/outlook.tex
\section{Outlook}
\label{sec:outlook}

\subsection{Generalizations of the Basic Architecture}

We presented one simple architecture (Figure~\ref{fig:arch}) which was motivated by the entropy and complexity of mean-field quantum models, especially the SYK model. The particular scaling-inspired ansatz with $k=r^a$ and $N = r^a + r^{a+b}$ is one instance of that architecture, but one could well imagine other choices.

Moreover, inspired by branching MERA~\cite{evenbly_class_highly_entangled_2014} and s-sourcery~\cite{swingle_renormalization_group_constructions_2016}, one can consider other architectures in which the added thermal degrees of freedom are not just in a product state. As a basic example, consider the following structure. Take the encoding circuits for two $[[n,k,\delta]]$ codes and mix their physical qubits using an additional depth $D$ quantum circuit. The result is a code on $n' = 2n$ qubits with $k'=2k$. Hence, the rate is the same, $k'/n' = k/n$. The distance will also increase by a factor with some probability, as will the weights of the stabilizers. Starting with a root $[[n_0,k_0,\delta_0]]$ code, $L$ layers of this construction produces a code with parameters $[[2^L n_0, 2^L k_0, \delta]]$ with $\delta$ some $L$-dependent distance.

In the above construction, the rate of the final code is determined by the rate of the root code. We could also vary the rate by introducing additional product qubits in each iteration of the process (analogous to the thermal qubits above) or by combining codes with different rates at each iteration. 

In all these constructions, the distance is also expected to grow exponentially with $L$, the number of layers. However, the weight of the checks will also generically grow when we use random depth-$D$ circuits. From this perspective, the challenge of producing a good quantum LPDC code is the challenge of keeping the weights of checks low while keeping the distance high. This clearly requires tuning of the layer circuits, likely made possible by the addition of some structure to the problem. In light of the recent rapid progress in the area of good quantum LDPC codes, it would be interesting to understand if our architecture can capture these recently discovered codes.


\subsection{Further Weight Reductions}
\label{sec:weight_reduction}

One direction we intend to explore in the future is finding alternative bases of given generated stabilizer code that minimize the overall weights. In the exact case, such a basis is unique and given by the \emph{reduced-row echelon form (RREF)} of the stabilizer matrix i.e.\ the matrix with all stabilizer basis elements as row vectors.
The RREF for a general matrix is defined in terms of the following rules:
\begin{enumerate}
    \item All rows consisting of only zeroes are at the bottom.
    \item The leading entry (i.e.\ the left-most non-zero entry) of every non-zero row is a 1, and to the right of the leading entry of every row above.
    \item Each column containing a leading 1 has zeros in all its other entries.
\end{enumerate}
In the case of a stabilizer matrix the first rule can be ignored since the matrix has maximum rank. The remaining two requirements can be easily met by applying a Gaussian elimination algorithm. An example of a possible resulting RREF is given by
\begin{equation}
    \begin{pmatrix}
        1 & 0 & a_1 & 0 & b_1 & 0 \\
        0 & 1 & a_2 & 0 & b_2 & 0 \\
        0 & 0 & 0 & 1 & b_3 & 0 \\
        0 & 0 & 0 & 0 & 0 & 1
    \end{pmatrix}
\end{equation}
From this it should be intuitively clear that the RREF maximizes the number of zero-valued elements in the matrix, thus minimizing the weights of the stabilizer basis elements.

Another possible method of weight reduction could be finding a set of stabilizers that approximately replicates our model, but whose basis has low weights. A potential way to achieve this is to use so-called \emph{perturbative gadgets} \cite{jordan_perturbative_gadgets_arbitrary_2008}. Considering the Hamiltonian representation \eqref{eq:stab_hamiltonian} of a given stabilizer code $[[n,k,\delta]]$, each term in the sum acts on a number of qudits equivalent to its weight. Let $w$ be the largest weight of all terms in the Hamiltonian, then we speak of a $w$-local Hamiltonian. Using $w$th-order perturbation theory it can then be shown that there must exist a 2-local Hamiltonian that approximately has the same ground space i.e.\ the desired quantum error-correcting code. Said Hamiltonian is called the \emph{gadget Hamiltonian} and its construction involves introducing $w \cdot (n-k)$ ancillary qudits, where $n-k$ is the number of terms in the original Hamiltonian. The approximate ground space Hamiltonian can then be recovered by block-diagonalizing the gadget Hamiltonian and only considering the entries with unit eigenvalue in the ancillary space. It should be noted that the terms of this resulting Hamiltonian might not necessarily commute anymore, but we expect the ground state degeneracy and code properties to be unaffected. 
In future work we intend to explore both approaches for weight reduction in the context of our tensor network ansatz and compare them to the baseline considerations made in this paper.

\subsection{Towards a Closer Link With SYK}

It is also interesting to move towards closer contact with SYK. The first step is to develop a fermionic analogue of our architecture. Then, because the network is not efficiently contractible in general on a classical computer, it is interesting to pursue a quantum simulation strategy where we treat our architectecture as a variational ansatz. The variational parameters would be the gates within each circuit layer as well as the discrete data of the network, e.g. the number of qubits at each layer. It would also be important to understand and adapt the construction to some of the details of SYK, e.g. the specific expected form of the ground state degeneracy.

A related setting where we should be able to carry out classical simulations is the SYK model for $q=2$, i.e. a non-interacting fermion model with random all-to-all hoppings. In this case, we can efficiently simulate the network using non-interacting fermion machinery. There is no ground state degeneracy for the random hopping model, so $k=0$ in this case, but one could still test other properties of the network. We are currently exploring this direction.

In the spirit of generalizing to fermionic models, it is also interesting to consider fermionic generalizations of the Clifford formalism, e.g. the subgroup of the full set of fermionic unitaries that maps strings of fermion operators to other strings of fermion operators. By developing methods to sample these transformations and to compute entropies of subsets of the fermions, one would be able to repeat the studies in this work in the language of fermionic codes~\cite{bravyi_majorana_fermion_codes_2010}. This is also a work in progress.

\subsection{NoRA as an Ansatz for Approximating Mean-Field Ground States}

In addition to the SYK model, there are numerous other mean-field models whose ground states we might try to model with a NoRA network. One large class consists of quantum spin glass models. Here we explain with an example why NoRA is a plausible ansatz for this class of models. 

For concreteness, consider a quantum transverse field Sherrington-Kirkpatrick (TFSK) model on $i=1,\cdots,N$ qubits,
\begin{equation}
    H = \sum_{i<j} J_{ij} \sigma_i^z \sigma^z_j + g \sum_i \sigma^x_i.
\end{equation}
The couplings $J_{ij}$ are Gaussian random variables with zero mean and variance $1/N$. This model is known to have a quantum phase transition between a non-glassy state at large $g$ and a glassy state at small $g$. At $g=\infty$, the ground state is simply a product state $\sigma^x$ eigenstates, while at $g=0$ the model reduces to the classical Sherrington-Kirkpatrick spin glass.

At $g=\infty$, the ground state is a product state and hence trivially a NORA network with zero layers. The energy gap between the ground state and the first excited state is non-vanishing at large $N$. At large but finite $g$, the model is still in the non-glassy phase and the gap between the ground state and the first excited state remains non-vanishing at large $N$. Because of the non-vanishing gap, we can presumably approximately prepare the ground state at large but finite $g$ using an adiabatic evolution for a constant time proportional to $1/g$ (since the gap is proportional to $g$). This is analogous to one layer of the NORA network. Hence, it is plausibly the case that a NoRA with one layer could capture the ground state of the TFSK using a NoRA network (in fact, one which requires just one layer as opposed to $\ln N$ layers).

Similarly, at $g=0$ the model reduces to the classical Sherrington-Kirkpatrick spin glass and the ground state is a product state. Deep in the glassy phase at small but non-zero $g$, the ground state is also caricatured by a product state and it is also plausible that a single-layer NoRA can capture the ground state.

Finally, at the critical point separating the two phases, it is less clear whether NoRA is suitable or not, but there is still a scaling symmetry present in the model and so NoRA wth the scaling ansatz and $k=0$ is still a reasonable candidate for describing the ground state wavefunction. 

We emphasize that these are plausibility arguments. We hope to carry out a more systematic study of this direction in future work. These plausibility arguments can also be adapted to a variety of other mean-field models. For example, other models with gapless critical points or critical phases may have a NORA-like description which requires the full layer structure with $\ln N$ layers. We have also conjectured that general quantum LDPC codes may have a NORA-like description and in such cases presumably the full layer structure is also needed, i.e. a finite depth circuit, even without geometric locality constraints, would not be sufficient to capture the ground space. These are all interesting directions for future work, especially the question of what determines the required number of layers in a NORA network.

\subsection{Connections With Holographic Models}

Finally, let us comment further on the connection to low-dimensional models of quantum gravity. We already made use of these connections as part of the motivation for our ansatz, in particular, we checked the complexity of our network against holographic estimates of complexity.

The basic point is that our architecture mimics the structure of two-dimensional Jackiw-Teitelboim (JT) gravity in AdS2. The analogue of the area formula for black hole entropy is the statement that the entropy of black hole is equal to the value of a scalar field called the dilaton evaluated at the event horizon (technically, at the bifurcation point), see e.g.~\cite{jafferis_entanglement_entropy_jackiw_2022}. This in turn leads to a low-dimensional version of the Ryu-Takayanagi formula for entanglement (for a review see \cite{rangamani_holographic_entanglement_entropy_2017}), also in terms of the dilaton field.

After solving the equations of motion, one finds a solution in which the metric is 
\begin{equation}
    ds^2 = \frac{-dt^2 + dz^2}{z^2}
\end{equation}
and the dilaton is
\begin{equation}
    \phi(z) = \phi_1/z.
\end{equation}
The number of degrees of freedom at a scale determined by $z$ is proportional to
\begin{equation}
    \phi_0 + \phi(z),
\end{equation}
where $\phi_0$ is some background value. Comparing to our network, we interpret $\ln z$ as analogous to $L-\ell$, which increases from zero as we go from the UV (top of Fig.~\ref{fig:arch}) to the IR (bottom of Fig.~\ref{fig:arch}) . The number of qudits at ``layer $z$'' would be $k+ (N-k) e^{- \ln z}$. Then, since $\phi_0 + \phi(z)$ also has the form
\begin{equation}
    \text{constant} + \text{constant}' e^{- \ln z},
\end{equation}
we could interpret $\frac{\phi_0}{4 G_N}$ as analogous to $k$ and $\frac{\phi_1}{4 G_N}$ as analogous to $N-k$. 

\begin{figure}
    \centering
    \includegraphics[width=\linewidth]{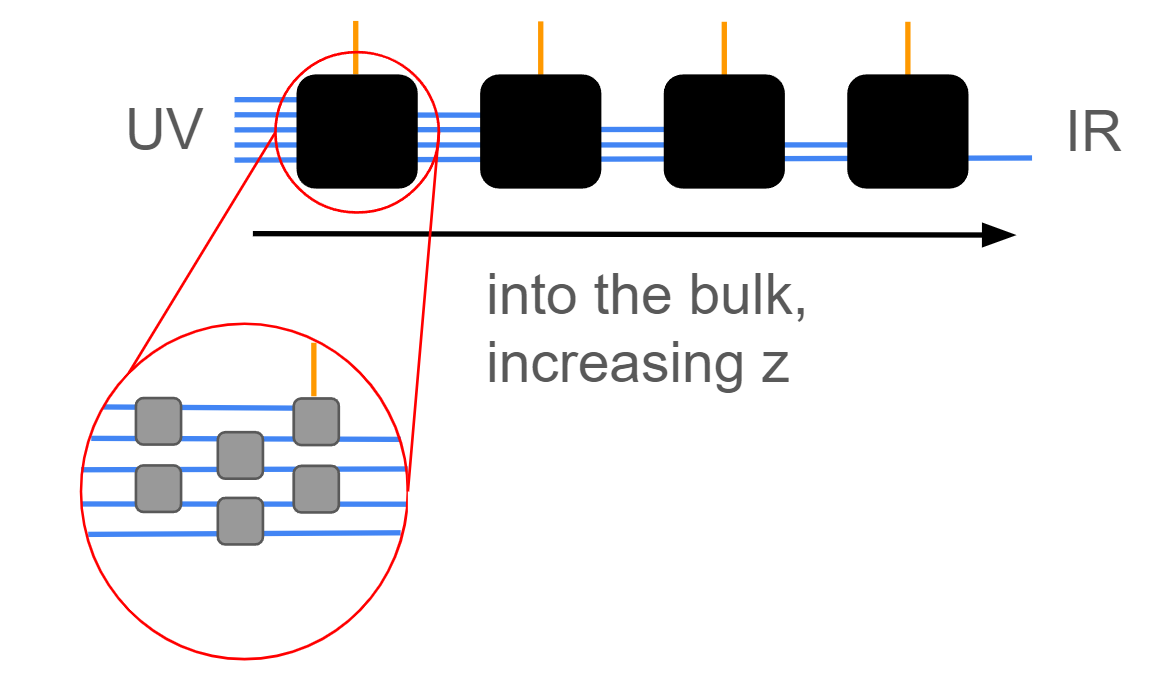}
    \caption{Schematic of a matrix product state as simple toy model of a spatial slice of a $1+1$d holographic geometry. The orange upward lines can be viewed as bulk degrees of freedom and are analogous to the $|0\rangle$ input states in NoRA (see Figure~\ref{fig:arch}). The shrinking number of blue lines as one proceeds into the bulk provides a schematic version of the decrease of degrees of freedom discussed just above. The NoRA network can be viewed as a refinement of this simple matrix product state model in which the tensors of the matrix product state have additional substructure. Here the orientation of the diagram is rotated $90$ degrees relative to Figure~\ref{fig:arch}.}
    \label{fig:holo_mps}
\end{figure}

It is also instructive to compare the tensor network model we just described with a less structured model. Let us step back and start by just supposing that the tensor network is a model of the spatial geometry. For two or more spatial dimensions in the bulk, this can give an interesting network structure. For example, the HaPPY code and its associated network yields a discretization of the two-dimensional hyperbolic disk. However, in one dimension, there is no interesting geometry and the analogue of the HaPPY network would be a simple one-dimensional network, essentially a matrix product state. We could identify the direction along the network with $\ln z$ and let the bond dimension of the network depend on $z$ such that $\ln \chi(z) \propto \phi_0 + \phi(z)$. But apart from these choices, the model is unstructured and nothing yet has been said about the structure of the tensors in the matrix product state. Within this framing, we can view the NoRA network model as a refinement of the matrix product state model in which we endow the tensors with the layer structure of the NoRA network as in Figure~\ref{fig:holo_mps}.

\paragraph{Acknowledgements}

We thank Isaac Kim, Vincent Su, Michael Walter and Gregory Bentsen for discussions. We acknowledge support from the U.S. Department of Energy grant DE-SC0009986 (V.B.) and from the AFOSR under grant number FA9550-19-1-0360 (B.G.S.).

%% file: glossary.tex
\section*{Glossary}

\begin{tabularx}{\columnwidth}{r X}
    $d$ & Local qudit bond dimension. \\
    $L$ & Number of layers in the NoRA network. \\
    $n_{\ell}$ & Number of qudits being acted on non-trivially at layer $\ell$. \\
    $N$ & Number of physical qudits. Equal to $n_L$. \\
    $k$ & Number of (ground-state) logical qudits. Equal to $n_0$. \\
    $\Delta n_{\ell}$ & Number of new (thermal) qudits introduced at layer $\ell$. Equal to $n_{\ell} - n_{\ell - 1}$. \\
    $D$ & Circuit depth for a single layer of NoRA (layer-independent). \\
    $r$ & Qudit growth rate of the NoRA network with specified scaling. \\
    $q$ & Locality of the circuit at a single later (layer-independent).  \\
    $w_{\ell}$ & Operator weight of a (stabilizer) Pauli string at layer $\ell$. \\
    $g$ & Effective weight growth rate for a single random $q$-local Clifford. Approximately $q \cdot (d^2 - 1)/d^2$. \\
    $D_{\text{sat}}$ & Minimum layer circuit depth necessary to achieve approximate weight saturation. Approximately $ \approx \log_g r$. \\
    $\delta$ & Code distance of a $[[N, k, \delta]]$ error-correcting code. \\
    $\delta_{\text{qsb}}$ & Maximum possible code distance as predicted by the quantum singleton bound. Equal to $(N - k)/2 + 1$. \\
    $\delta / N$ & Relative code distance.
    
\end{tabularx}

%% file: appendices/entropy_scaling.tex
\section{Scaling of the Stabilizer Entropy}
\label{sec:entropy_scaling}

\subsection{The Stabilizer Hamiltonian}

Every set of stabilizers fixing a quantum state or a space of quantum states can be expressed as a projective Hamiltonian that has said states as part of its (degenerate) ground space.

Let $N$ be the total number of physical qudits and $k \leq N$ the number of logical qudits needed to represent the code word $\ket{\psi}_{\textrm{code}}$. The case of $N = k$ is not interesting to us so we assume we have $N - k > 0$ ancillary qudits. Every quantum code can then be written as a unitary $U$ satisfying
\begin{equation}
    \ket{\Psi} = U \left( \ket{\psi}_{\textrm{code}} \otimes \ket{0}_{\textrm{anc}}^{\otimes \, (N - k)} \right),
\end{equation}
where $\ket{\Psi}$ is the code word encoded in the space of physical qudits. The ancillary thermal qudits can each be fixed to be $\ket{0}$ without loss of generality.

Before applying the encoding unitary, it is easy to see that the generating set of stabilizers fixing the code space spanned by all possible choices of $\ket{\psi}_{\textrm{code}} \otimes \ket{0}_{\textrm{anc}}^{\otimes \, (N-k)}$ is given by
\begin{equation}
\label{eq:pre_code_stab}
    Z_i \equiv I_{\textrm{code}} \otimes I^{\otimes \, (i-1)} \otimes Z \otimes  I^{\otimes \, (N-k-i)}, \quad i = 1,\ldots,N-k, 
\end{equation}
where the $Z$ acts on the $i$th qudit of the ancillary system. From this it immediately follows that the stabilizers acting on the physical qudits can be retrieved by applying the code unitary such that
\begin{equation}
    \widetilde{Z}_i = U \, Z_i \, U^{\dagger}.
\end{equation}

To construct the stabilizer Hamiltonian though, we have to use the (disjoint) projectors associated to our chosen stabilizer basis. Analogously to the previous case, before encoding the state they are
\begin{equation}
\label{eq:pre_code_proj}
    P_i \equiv I_{\textrm{code}} \otimes I^{\otimes \, (i-1)} \otimes \ket{0}\bra{0} \otimes I^{\otimes \, (N-k-i)}, \quad i = 1,\ldots,N-k, 
\end{equation}
and after the encoding they become
\begin{equation}
    \widetilde{P}_i = U \, P_i \, U^{\dagger}.
\end{equation}

With that, the general stabilizer Hamiltonian has the form of
\begin{equation}
\label{eq:stab_hamil}
    H = - \sum_{i = 1}^{N-k} J_i \cdot \widetilde{P}_i, \quad J_i > 0
\end{equation}
The coefficients $J_i$ can be arbitrarily chosen and determine the energy scales of the system, but since they are necessarily positive-definite, this do not affect the space of ground states i.e. the space of valid physical qudit states. Excitations away from a ground states then correspond to errors being present in the state, which is because of the one-to-one relation between projectors and stabilizers.

\subsection{General Thermodynamic Quantities}

Using the Hamiltonian derived in the previous section, we can now compute the associated Gibbs state
\begin{equation}
    \rho_{\beta} = \frac{1}{Z} e^{-\beta H}, \quad Z = \Tr[e^{-\beta H}]
\end{equation}
and some of its properties, including the entropy.

First, it is straightforward to show that
\begin{align}
\begin{split}
\label{eq:gibbs_exp}
    e^{-\beta H} &= \exp\left( \beta \sum_{i = 1}^{N-k} J_i \cdot \widetilde{P}_i \right) \\
    &= \prod_{i = 1}^{N-k} \exp\left(\beta J_i \cdot \widetilde{P}_i\right) \\
    &= \prod_{i = 1}^{N-k} \left[ \sum_{n=0}^{\infty} \frac{1}{n!} \left(\beta J_i \cdot \widetilde{P}_i \right)^n \right] \\
    &= \prod_{i = 1}^{N-k} \left[ I + \sum_{n=1}^{\infty} \frac{1}{n!} \left(\beta J_i\right)^n \cdot \widetilde{P}_i\right] \\
    &= \prod_{i = 1}^{N-k} \left[ I + \left(e^{\beta J_i} - 1 \right) \cdot \widetilde{P}_i\right] \\
    &= \sum_{n=0}^{N-k} \, \sum_{1 \leq i_1 < \ldots < i_n \leq N-k} \, \prod_{\{i_a\}} \left( e^{ \beta J_{i_a}} - 1\right)\cdot \widetilde{P}_{i_a},
\end{split}
\end{align}
where in the last line we used a generalization of the binomial theorem and the fact that the projection operators commute by definition. Computing the partition function $Z$ using the final expression in \eqref{eq:gibbs_exp} can be done in the following way:
\begin{align}
\begin{split}
    Z &= \Tr[e^{-\beta H}] \\
    &= \sum_{n=0}^{N-k} \, \sum_{1 \leq i_1 < \ldots < i_n \leq N-k} \, \prod_{\{i_a\}} \left( e^{ \beta J_{i_a}} - 1 \right) \cdot \Tr \bigg[ \prod_{\{i_a\}}\widetilde{P}_{i_a} \bigg] \\
    &= \sum_{n=0}^{N-k} \, \sum_{1 \leq i_1 < \ldots < i_n \leq N-k} d^{N-n} \cdot \prod_{\{i_a\}} \left( e^{ \beta J_{i_a}} - 1 \right) \\
    &= d^{k} \cdot \sum_{n=0}^{N-k} \, \sum_{1 \leq i_1 < \ldots < i_n \leq N-k} d^{N-k-n} \cdot \prod_{\{i_a\}} \left( e^{ \beta J_{i_a}} - 1 \right) \\
    &= d^k \cdot \prod_{i=1}^{N-k} \left( e^{ \beta J_i} + d - 1 \right).
\end{split}
x\end{align}
Note that in the second line we used the definition \eqref{eq:pre_code_proj} for the projection operators, which implies that $\Tr[\widetilde{P}_{i_1} \cdots \widetilde{P}_{i_n}] = d^{N-n}$ given that none of the indices $i_a$ coincide. Going from the penultimate line to the last one we then again applied the generalized binomial theorem.

From the partition function it is then easy to determine all other thermodynamic quantities, of which the most important one for us is the von-Neumann entropy
\begin{align}
\begin{split}
    S &\equiv - \Tr[\rho_{\beta} \log(\rho_{\beta})] \\
    &= \beta \cdot \braket{E}_{\beta} + \log(Z) \\
    &= (1 - \beta \cdot \partial_{\beta}) \log(Z),
\end{split}
\end{align}
where the second and third lines are well-known equivalent expressions and we assume $\log$ to refer to the natural logarithm. Therefore, by using the fact that
\begin{align}
    \log(Z) &= k \cdot \log(d) + \sum_{i=1}^{N-k} \log\left( e^{ \beta J_i} + d - 1 \right), \\
    - \beta \cdot \partial_{\beta} \log(Z) &= - \sum_{i=1}^{N-k} \frac{\beta J_i \cdot e^{\beta J_i}}{e^{\beta J_i} + d - 1},
\end{align}
and after doing some rearranging, we arrive at
\begin{equation}
    S_{\textrm{stab}} = \left( k \log(d) + \sum_{i=1}^{N-k} p_i \log(d-1) \right) + \sum_{i=1}^{N-k} S(p_i),
\end{equation}
where
\begin{equation}
    S(p_i) \equiv - p_i \cdot \log(p_i)) - (1-p_i) \cdot \log(1-p_i)
\end{equation}
is the binary Shannon entropy associated to the probability distributions $\{p_i, 1-p_i\}_i$ which are defined in terms of
\begin{equation}
\label{eq:shannon_prob}
    p_i \equiv \frac{d-1}{e^{\beta J_i} + d - 1} \in \left(0, \frac{d-1}{d}\right).
\end{equation}
Note that in the case of qubits ($d = 2$), $p_i$ is the Fermi-Dirac distribution associated to $J_i$. Hence we can interpret the sum in the leading term as an occupation number such that
\begin{equation}
    \braket{N-k} \equiv \sum_{i=1}^{N-k} p_i, \quad S_{\textrm{stab}} = \log\left(d^k \cdot (d-1)^{\braket{N-k}}\right) + \sum_{i=1}^{N-k} S(p_i).
\end{equation}
Ignoring that leading term, the total entropy of the Gibbs ensemble therefore decouples into a sum of entropies associated with each energy level $J_i$ and therefore each element of the stabilizer basis \eqref{eq:pre_code_stab}. This is not unexpected though, as each term in the stabilizer Hamiltonian \eqref{eq:stab_hamil} commutes with every other one, making the system completely diagonalizable.

\subsection{Entropy Scaling for the NoRA Model}

So far all the calculations we did hold for error-correcting stabilizer codes in general. To actually get some results unique to the NoRA network discussed in this paper, we have to make some assumptions about the distribution of energy levels $J_i$.

One obvious such assumption is that the level distribution should only depend strongly on the layer $\ell$ at which associated stabilizer elements are first acted on in a non-trivial way by the encoding unitary. Hence we move from $J_i$ to $J_{\ell}$ (and therefore from $p_i$ to $p_{\ell}$),  ignoring (for now) that the energy might actually vary slightly for different stabilizers at the same level. Because of this the expression for the entropy becomes
\begin{equation}
\label{eq:level_entr_discrete}
    S_{\textrm{stab}} = \log\left(d^k \cdot (d-1)^{\braket{N-k}}\right) + \sum_{\ell=1}^{L} \Delta n_{\ell} \cdot S(p_{\ell}),
\end{equation}
where $n_{\ell}$ is the number of stabilizer basis elements with the same associated energy level:
\begin{align}
\begin{split}
\label{eq:stab_distribution}
    \Delta n_{\ell=1} &= r, \\
    \Delta n_{\ell > 1} &= r^{\ell} - n_{\ell - 1} = (r - 1) \cdot r^{\ell - 1}.
\end{split}
\end{align}
with $1 \leq \ell \leq L$ and $r^L = N - k$. It is easy to see that this distribution therefore does indeed satisfy $\sum_{\ell} \Delta n_{\ell} = N - k$.

The other assumption we are making is that the distribution of energies $J_{\ell}$ increases exponentially with increasing $\ell$, giving it the form of
\begin{equation}
\label{eq:energy_distribution_disc}
    J_{\ell} = \Lambda \cdot e^{-\gamma \cdot (L - \ell)}
\end{equation}
for some UV energy scale $\Lambda > 0$ and rate of increase $\gamma > 0$. This is an artificial but reasonable choice because we want the circuit to obey renormalization invariance while going from the IR to UV limit in the same was as MERA networks generally do.

\subsubsection{Moving to the Continuum Limit}

To determine the scaling of the entropy close to the zero temperature (i.e. $\beta \rightarrow \infty$) limit, it is useful to consider the continuum limit of \eqref{eq:level_entr_discrete} in addition to the other assumptions we made. The stabilizer difference $\Delta n_{\ell}$ therefore becomes the stabilizer density
\begin{equation}
    \rho(\ell) = 
    \rho_0 \cdot e^{\alpha \cdot \ell}, \quad \ell \in [0, L],
\end{equation}
where $\alpha > 0$ can be chosen arbitrarily\footnote{One could of course choose $\alpha = \log(r)$ in the spirit of \eqref{eq:stab_distribution}, but we will refrain from making a specific choice here for the sake of generality. This specific case will be considered later when comparing the approximation with the actual entropy formula.} and $\rho_0$ is fixed by the density having to satisfy
\begin{equation}
    N-k \stackrel{!}{=} \int_0^L d\ell \, \rho(\ell) = \frac{\rho_0}{\alpha} \left( e^{\alpha \cdot L} - 1 \right) \quad \Longleftrightarrow \quad \rho_0 = \frac{\alpha \cdot (N-k)}{e^{\alpha \cdot L} - 1}.
\end{equation}
Because the distribution of the energy levels \eqref{eq:energy_distribution_disc} can be left untouched when moving to the continuum limit, the stabilizer entropy can be naively approximated as
\begin{equation}
\label{eq:level_entr_cont}
    S_{\textrm{stab}} \approx S_{\textrm{cont}} = \log\left(d^{k} \cdot (d-1)^{\braket{N-k}}\right) + \int_0^L d\ell \, \rho(\ell) \cdot S(p(\ell)),
\end{equation}
with $p(\ell)$ being of the same form as $p_{\ell}$ in \eqref{eq:shannon_prob}, but now considered as a continuous function of $\ell$. But to make the upcoming calculations easier, we perform a change of variables, integrating over $J = J(\ell)$ instead of $\ell$. To do that, we first note that from \eqref{eq:energy_distribution_disc} it follows that
\begin{equation}
    \ell(J) = L + \frac{1}{\gamma} \cdot \log\left( \frac{J}{\Lambda} \right),
\end{equation}
and hence
\begin{equation}
    d \ell = \frac{d \ell}{d J} \, dJ = \frac{dJ}{\gamma \cdot J} .
\end{equation}
This also allows us to express the stabilizer density as a function dependent on $J$:
\begin{equation}
    \rho(J) = \rho_0 \cdot e^{\alpha L} \cdot \left( \frac{J}{\Lambda} \right)^{\alpha/\gamma}.
\end{equation}
Finally, the continuous entropy as an integral over $J$ is
\begin{align}
\begin{split}
\label{eq:S_cont}
    S_{\textrm{cont}} &= \log\left(d^{k} \cdot (d-1)^{\braket{N-k}}\right) + \int_{
    \Lambda \cdot e^{-\gamma  L}}^{\Lambda} dJ \, \frac{\rho(J)}{\gamma \cdot J} \cdot S(p(J)) \\
    &= \log\left(d^{k} \cdot (d-1)^{\braket{N-k}}\right) + \frac{\rho_0 }{\gamma} \cdot e^{\alpha  L} \cdot \int_{
    \Lambda \cdot e^{-\gamma L}}^{\Lambda} \frac{dJ}{\Lambda}  \left( \frac{J}{\Lambda} \right)^{\alpha/\gamma - 1} \cdot S(p(J)).
\end{split}
\end{align}
Note that the lower integration bound acts as an effective IR cutoff for the integral. This is necessary for us to be able to make the following approximations..

\subsubsection{Low-Temperature Limit}

Computing the integral in \eqref{eq:S_cont} is in general hard, but since we are only interested in the limit of small $T/J$ (or equivalently large $\beta J$), we can approximate the binary entropy $S(p(J))$ that occurs in the integral as
\begin{align}
\begin{split}
\label{eq:binS_approx}
    S(p(J)) &= - \frac{d-1}{e^{\beta J} + d - 1} \cdot \log \left( \frac{d-1}{e^{\beta J} + d - 1} \right) -  \frac{e^{\beta J}}{e^{\beta J} + d - 1} \cdot \log \left( \frac{e^{\beta J}}{e^{\beta J} + d - 1} \right) \\ 
    &\stackrel{\beta J \rightarrow \infty}{=} (d-1) \cdot \frac{\beta J}{e^{\beta J}} + \mathcal{O}(e^{-\beta J}),
\end{split}
\end{align}
which is straightforward to prove. To realize this limit it is necessary to choose the right parameters since it follows from \eqref{eq:energy_distribution_disc} that
\begin{equation}
    \beta J = \beta \Lambda \cdot e^{- \gamma (L-\ell)} \gg 1 \quad \forall \, \ell
\end{equation}
and hence
\begin{equation}
    \beta \Lambda \cdot e^{-\gamma L} \gg 1 \quad \Longleftrightarrow \quad \gamma L \ll \log(\beta \Lambda).
\end{equation}

\begin{figure}[hbt]
    \centering
    \includegraphics{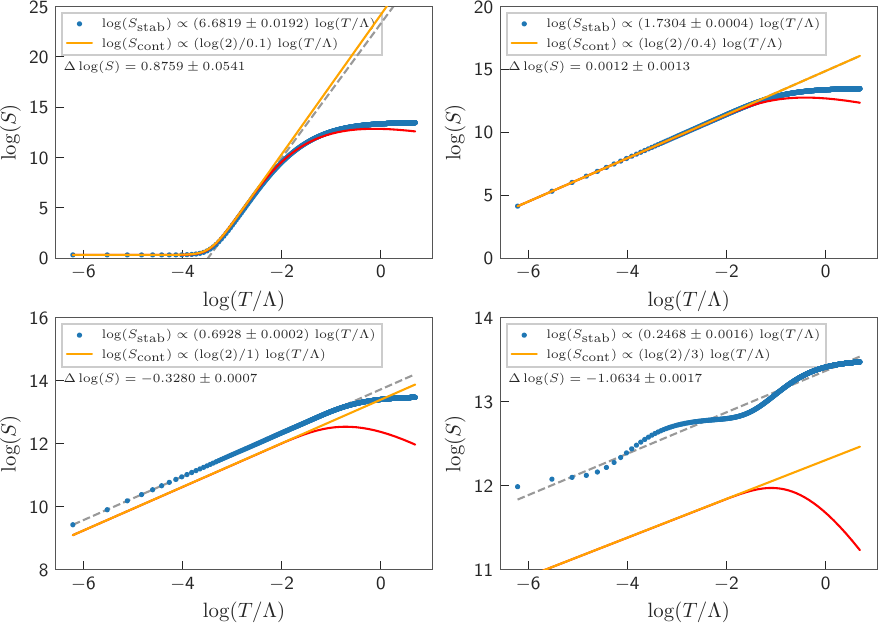}
    \caption{Logarithmic scaling of exact stabilizer entropies $S_{\textrm{stab}}$ and their continuous approximations $S_{\textrm{cont}}$ (with and without the gamma function correction) for $L=20$, $N-k=r^L$, $k=1$, $d=2$, $r=2$, $\alpha = \log(r)$, $\Lambda=1$ and $\gamma \in \{0.1, 0.4, 1, 3\}$. In the first two figures it can be seen that our continuous approximation from \eqref{eq:S_cont_2} matches almost exactly with the discrete stabilizer entropy for $\gamma \ll 1$ and small $T/\Lambda$. Even though the second figure shows less behavior than the first one, we expect that it will behave similarly for even lower relative temperatures. While the last two approximations with $\gamma \geq 0$ also receive their primary contribution from the polynomial term, it is more apparent that they don't completely align with the actual data anymore. Especially in the last figure where $\gamma = 2$ the trend of the stabilizer entropy is not strictly polynomial anymore. Still, each figure has at least a regime where its growth is either exactly polynomial or follows a polynomial trend that aligns with our theoretical predictions up to a total constant factor.}  
    \label{fig:entropy_scaling_appendix}
\end{figure}

Plugging \eqref{eq:binS_approx} into \eqref{eq:S_cont} and noting that $\braket{N-k} = \sum_i p_i = 0$ in that limit then leaves us with an expression that can be further simplified using a change of variables:
\begin{align}
\begin{split}
\label{eq:S_cont_2}
    S_{\textrm{cont}} &\approx k \log(d) + (d-1) \cdot \frac{\rho_0 \cdot e^{\alpha L}}{\gamma}  \int_{
    \Lambda \cdot e^{-\gamma L}}^{\Lambda} \frac{dJ}{\Lambda} \frac{\beta J}{e^{\beta J}} \left( \frac{J}{\Lambda} \right)^{\alpha/\gamma - 1} \\
    &= k \log(d) + (d-1) \cdot \frac{\rho_0 \cdot e^{\alpha L}}{\gamma} \cdot (\beta \Lambda)^{-\alpha/\gamma} \int_{\beta \Lambda \cdot e^{-\gamma L}}^{\beta \Lambda} dt \, t^{\alpha/\gamma} \cdot e^{-t} \\
    &= k \log(d) + (d-1)(N-k) \cdot \frac{\alpha}{\gamma} \cdot \frac{e^{\alpha L}}{e^{\alpha L} - 1} \cdot (\beta \Lambda)^{-\alpha/\gamma} \int_{\beta \Lambda \cdot e^{-\gamma L}}^{\beta \Lambda} dt \, t^{\alpha/\gamma} \cdot e^{-t} \\
    &\stackrel{\alpha L \gg 1}{\approx} k \log(d) + (d-1) (N-k) \cdot \frac{\alpha}{\gamma} \cdot (\beta \Lambda)^{-\alpha/\gamma} \int_{\beta \Lambda \cdot e^{-\gamma L}}^{\beta \Lambda} dt \, t^{\alpha/\gamma} \cdot e^{-t}
\end{split}
\end{align}
Let's consider the trailing integral. Up to the integration bounds it is the same as the gamma function $\Gamma(\alpha/\gamma + 1)$, whose integrand is positive everywhere. We can therefore get an upper bound for $S_{\textrm{cont}}$ (that we also expect to be approximately saturated for certain domains of $\beta \Lambda$) by substituting the \enquote{incomplete} gamma function with the proper one. Thus we have
\begin{equation}
\label{eq:S_cont_3}
    S_{\textrm{cont}} \lessapprox  k \log(d) + (d-1) (N-k) \cdot \frac{\alpha}{\gamma} \cdot \Gamma\left(\frac{\alpha}{\gamma} + 1\right) \cdot (\beta \Lambda)^{-\alpha/\gamma}, 
\end{equation}
which only scales with $(\beta \Lambda)^{-\alpha/\gamma} = (T/\Lambda)^{\alpha/\gamma}$, indicating that the entropy could indeed follow a power law, at least for certain low-temperature regimes. To show how well both continuous approximations hold up against the discrete stabilizer entropy with equivalent parameters ($N-k = r^L$, $\alpha=\log(r)$), we display both in logarithmic plots over $\log(T/\Lambda)$ and with different choices of $\gamma$, which is the only significant free parameter. These plots are depicted in figure \ref{fig:entropy_scaling_appendix} and indeed confirm that our low-temperature approximations are good at predicting aspects of the actual entropy, including its power-law growth.

%% file: appendices/growth_factor.tex
\section{Estimating the Layer Growth Factor}
\label{sec:growth_factor}

Given a generic string of $n$ generalized Pauli operators with local dimension $d$ and initial weight $w_0 \ll n$, we can estimate the relative weight growth $g$ the string experiences from one layer of $n/q$ random Cliffords being applied to random disjoint substrings of length $q$. The weight $w_k$ at the $k$th layer can therefore be estimated as
\begin{equation}
    w_k \approx g \cdot w_{k-1}.
\end{equation}

\subsection{Single Layer}

To find $g$ it is helpful to look at a single substring of length $q$ being scrambled by a single random Clifford. In that case, as long as a substring's weight $w_{k-1}(q)$ is not zero, we can expect its weight in the next layer to be on average
\begin{equation}
    \label{eq:substring_weight}
    w_k(q) = q \cdot \frac{d^2-1}{d^2},
\end{equation}
regardless of how the initial string looked\footnote{Remember that the generalized Pauli group of dimension $d$ has $d^2$ different elements, up to phases. Of those only the identity has zero weight.}. To extend this argument to the whole Pauli string we can therefore distinguish between two extreme cases:
\begin{itemize}
    \item All $w_{k-1}$ non-trivial Pauli operators are contained in as few substrings as possible, namely $\lceil w_{k-1}/q \rceil$. Since each such substring will on average have the weight \eqref{eq:substring_weight} after the Clifford layer, we find that
    \begin{equation}
    \label{eq:growth_saturated}
        w_k = \lceil w_{k-1}/q \rceil \cdot q \cdot \frac{d^2-1}{d^2} \approx w_{k-1} \cdot \frac{d^2-1}{d^2}.
    \end{equation}
    \item Each (non-trivial) substring contains exactly one non-trivial Pauli operator, meaning that in the next layer we have $w_{k-1}$ substrings each having the average weight \eqref{eq:substring_weight}. The total weight is therefore
    \begin{equation}
        w_k = w_{k-1} \cdot q \cdot \frac{d^2-1}{d^2}
    \end{equation}
\end{itemize}
Hence we can provide approximate upper and lower bounds for $g$ by
\begin{equation}
    \frac{d^2-1}{d^2} \lessapprox g \lessapprox q \cdot \frac{d^2-1}{d^2}.
\end{equation}
However, for our purposes we will always have $w_{k-1} \ll n$ (see next section), which makes it more likely for $g$ to be closer to the upper bound. Hence we can assume that
\begin{equation}
\label{eq:growth_factor}
    g \approx q \cdot \frac{d^2-1}{d^2}.
\end{equation}

\subsection{Multiple Layers}
\label{sec:max_depth}

Usually the scrambling circuits will be composed of more than one layer of random $q$-party Clifford gates. Therefore, given that we start with $w_0 \ll n$ and keep $g$ fixed to be \eqref{eq:growth_factor}, what is the approximate maximum depth $D$ for which $w_D = g^D \cdot w_0$ gives a good estimate for the total operator weight at the end?

\begin{figure}[htb]
    \centering
    \includegraphics{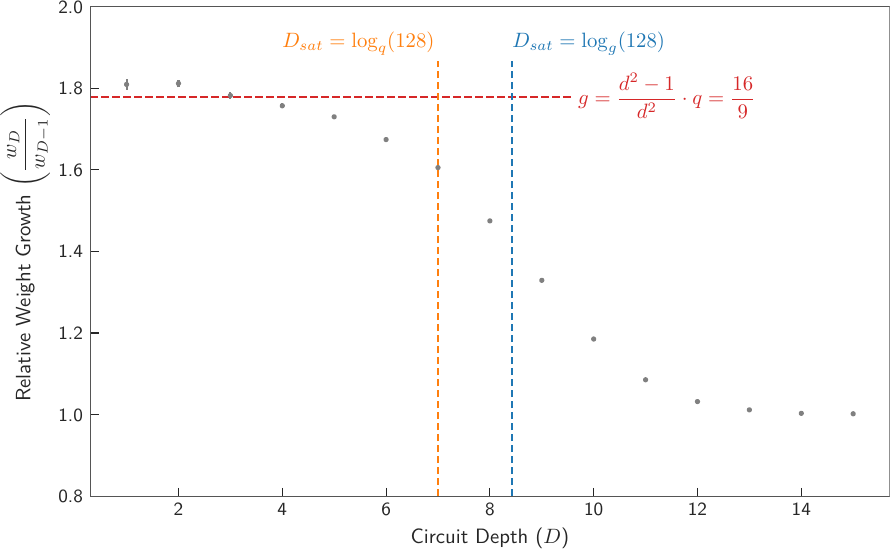}
    \caption{Averaged relative weight growth $w_D / w_{D-1}$ of a single Pauli string ($d=3, n=128, w_0 = 1$) subjected to a random 2-local Clifford with increasing circuit depth $D$ (1000 repetitions). Depicted are also the estimates of the effect growth factor $g$ \eqref{eq:growth_factor} and the saturation depth(s) $D_{\textrm{sat}}$ \eqref{eq:D_max} for which it can be considered to hold. While \eqref{eq:D_max} indeed provides a good maximum circuit depth for which the data and our estimate for $g = 16/9$ approximately coincide, a tighter bound can be achieved by instead using $q = 2$ as the base of the logarithm.}
    \label{fig:growth_factor}
\end{figure}

Due to our previous arguments, we can expect the approximation to not hold anymore by the time $w_D$ is of order $n$ since then the case of \eqref{eq:growth_saturated} will dominate. In this case we say that the weight is \emph{saturated}, and we can estimate the order of magnitude of the saturation depth $D_{\textrm{sat}}$ by requiring that $g^{D_{\textrm{sat}}} \cdot w_0 \lessapprox n$, leading to:
\begin{equation}
\label{eq:D_max}
    D_{\textrm{sat}} \approx \log_g \left( \frac{n}{w_0} \right).
\end{equation}
A tighter bound can also be achieved by using $\log_q$ instead of $\log_g$. Both options are shown for a specific simulated example in Figure \ref{fig:growth_factor}.

%% file: appendices/phase_space.tex
\section{Phase Space Formalism}
\label{sec:phase_space}

\subsection{Weyl Representation}
\label{sec:weyl_rep}

Given a Hilbert space $\mathcal{H}$ of prime dimension $d > 2$ \footnote{The case of $d=2$ is excluded here since our choice of representation requires the existence of a 2-element in the group such that $\frac{1}{2} \equiv \frac{d + 1}{2}$ is also a group element, which is not true for $d=2$ (i.e.\ the field cannot have characteristic 2).}, we choose a basis $\{\ket{0},\ket{1}, \ldots, \ket{d-1}\}$ with its states being labeled by the elements of the associated finite (Galois) field GF($d$) $\equiv \mathbb{F}_d$\footnote{Finite fields also exist for powers of primes i.e.\ GF($d^k$), but addition and multiplication does not happen mod $d^k$ then.}.  One can then introduce \emph{clock and shift operators} $Z, X$ which act on the basis states according to \cite{gross_hudson_theorem_finite_2006}
\begin{equation}
\label{eq:boost_shift}
    Z^p \ket{k} = \chi(p \cdot k) \ket{k}, \quad X^q \ket{k} = \ket{k + q},
\end{equation}
where $p, q, k \in \mathbb{F}_d$ and $\chi(k) = e^{2 \pi i k / d}$. Note that addition and multiplication happens over $\mathbb{F}_d$ and is thus mod $d$. This is also respected by our choice for $\chi(k)$ since $\chi(k + d) = \chi(k)$ even for addition without modulo.

We are now able to define the so-called \emph{Weyl operators} for a single qudit, which provide a generalisation of the Pauli operators on a qubit:
\begin{equation}
\label{eq:weyl_single}
    w(p, q) = \chi\left(-\frac{p \cdot q}{2}\right) \, Z^p \, X^q, 
    \quad p,q \in \mathbb{F}_d.
\end{equation}
Extending this definition to $n$ qudits is as easy as tensoring $n$ copies of \eqref{eq:weyl_single}, which we write as
\begin{align}
\label{eq:weyl_multi}
    \begin{split}
        w(v) &= w(p_1, q_1, \ldots, p_n, q_n) \\
        &= w(p_1, q_1) \otimes \ldots \otimes w(p_n, q_n).
\end{split}
\end{align}
Each Weyl operator is therefore uniquely represented by an element $v$ of a $2n$-dimensional vector space $V$ over the field $\mathbb{F}_d$. Using the commutation relations of $Z^p$ and $X^q$ that arise from their definition in \eqref{eq:boost_shift}, it also follows that
\begin{equation}
\label{eq:weyl_mul}
    w(v) \, w(w) = \chi \left( \frac{\symp{v}{w}}{2} \right) \, w(v + w),
\end{equation}
where $\symp{\cdot}{\cdot}$ is the \emph{symplectic product} on $V$, which obeys $\symp{v}{w} = -\symp{w}{v}$ and can be expressed as a matrix product:
\begin{equation}
\label{eq:symp_prod}
    \symp{v}{w} = v^T J w, \quad J = \begin{pmatrix}
        0 & 1 \\ -1 & 0
    \end{pmatrix}^{\oplus n}.
\end{equation}
Because of that the Weyl operators form a projective representation of the associated vector space $V$ equipped with a symplectic product. It is also noteworthy that \eqref{eq:weyl_mul} implies that two Weyl operators $w(v), w(w)$ commute if and only if the corresponding symplectic product $\symp{v}{w}$ vanishes.

Another useful identity which we will use later is the fact that
only the identity $I_n = w(0)$ has a non-vanishing trace:
\begin{equation}
    \label{eq:weyl_trace}
        \Tr[w(v)] = d^n \delta_{v,0}.
\end{equation}
This is trivial to show for $X^q$ but requires using the fact that the Kronecker delta can be written as
\begin{equation}
\label{eq:kronecker_sum}
    \delta_{p,0} = \frac{1}{d} \sum_{k = 0}^{d-1} e^{\frac{2 \pi i k}{d} p}
\end{equation}
to prove it for $Z^p$ as well. 

\subsection{The Clifford Group}

The Clifford group is a subset of the unitary group which maps Weyl operators to other Weyl operators (up to a factor):
\begin{equation}
\label{eq:clifford_def}
    U w(v) U^{\dagger} = c(v) \, w(S(v)),
\end{equation}
for some $c: V \rightarrow \mathbb{C}$ and $S: V \rightarrow V$. Because $S$ therefore has to be compatible with \eqref{eq:weyl_mul}, it is easy to see that it has to be linear and preserve the symplectic product:
\begin{equation}
    \symp{S v}{S w} = \symp{v}{w} \quad \forall \, v,w \in V.
\end{equation}
In matrix representation, one can also equivalently state this property as $S^T J S = J$. Such a function is called \emph{symplectic}. The set of all symplectic functions for a given vector space $V$ forms the so-called \emph{symplectic group} $\text{Sp}(2n, \mathbb{F}_d)$\footnote{Note the similarities to the definition of the orthogonal group. In fact, the column entries of a symplectic matrix also form as (symplectic) basis $(e_1, f_1, \ldots, e_n, f_n)$ of $V$ which satisfies $\symp{e_i}{e_j} = 0 = \symp{f_i}{f_j}$ and $\symp{e_i}{f_j} = \delta_{ij}$ for all $i,j = 1, \ldots, n$. Applying a symplectic is therefore equivalent to a change of basis. \label{fn:symp_basis}}. 

In general, the structure of the Clifford group is completely determined by the following statements:
\begin{enumerate}
    \item For any symplectic $S$ there is a unitary operator $\mu(S)$ satisfying
    \begin{equation}
        \mu(S) w(v) \mu(S)^{\dagger} = w(S v) \quad \forall \, v \in V.
    \end{equation}
    \item $\mu(S)$ is a projective representation of the symplectic group, meaning
    \begin{equation}
        \mu(S) \mu(T) = e^{i \phi} \mu(S T)
    \end{equation}
    for some phase $\phi$.
    \item Up to a phase, any Clifford operator is of the form
    \begin{equation}
        U = w(a) \mu(S)
    \end{equation}
    for a suitable $a \in V$ and symplectic $S$.
\end{enumerate}
A proof of these statements can be found in \cite{gross_hudson_theorem_finite_2006}. Note that this also fixes the factor from \eqref{eq:clifford_def} to be $c(v) = \chi(\symp{a}{Sv})$. 

\subsection{Stabilizer States and Codes}

As mentioned before, a vanishing symplectic product $\symp{v}{w}$ is equivalent to a vanishing commutator $[w(v), w(w)]$. One can therefore construct a set
\begin{equation}
    w(M) = \{ m \,|\, m \in M\}
\end{equation}
containing only commuting Weyl operators by choosing $M$ to be a subspace of $V$ satisfying
\begin{equation}
    \symp{m_i}{m_j} = 0 \quad \forall \, m_i, m_j \in M
\end{equation}
Such a subspace is called \emph{isotropic} and it is easy to see that it also forms a group under vector addition since the symplectic product is bilinear. The cardinality of isotropic subspaces can range between 0 and $d^n$ as there are at most $n$ elements with mutually vanishing symplectic product in a $2n$-dimensional symplectic basis (see footnote \ref{fn:symp_basis} for the reason). We will refer to $M$ having maximal cardinality as \emph{maximally isotropic}.

In general it is convenient to write the basis elements of an isotropic subspace as a $k \times 2n$ (or $2n \times k$) matrix over $\mathbb{F}_d$, where $k = \log_d(M)$ is the size of the basis. In the literature this is called the \emph{stabilizer matrix}, although there it is often written in terms of the actual Pauli/Weyl operators and not their symplectic representation.

Isotropy of $M$ allows one to (at least partially) diagonalize the Weyl operators contained in $w(M)$, even completely if $M$ is maximally isotropic. In the latter case it is therefore possible to define a unique quantum state $\ket{M, v}$ in terms of the elements in $w(M)$ acting on it as stabilizers:
\begin{equation}
\label{eq:stab_def}
    \chi(\symp{v}{m}) w(m) \ket{M, v} = \ket{M, v} \quad \forall \, m \in M.
\end{equation}
The vector $v \in V$ therefore determines the phase differences between the eigenstates assocated with $w(M)$. A state satisfying \eqref{eq:stab_def} is called a \emph{stabilizer state} and can be written as
\begin{equation}
\label{eq:stab_state}
    \ket{M, v}\bra{M, v} = \frac{1}{d^n} \sum_{m \in M} \chi(\symp{v}{m})\, w(m).
\end{equation}
It is easy to show that \eqref{eq:stab_state} is a projection operator and has unit trace by applying \eqref{eq:weyl_trace} and using the fact that $M$ is a group and thus satisfies $M + m = M$ for all $m \in M$.

In fact, even for a non-maximally isotropic subspace $M$ would \eqref{eq:stab_state} still be a projector (up to normalization), but not a quantum state anymore. In this more general case we have
\begin{equation}
\label{eq:stab_proj}
    \Pi(M,v) = \frac{1}{|M|} \sum_{m \in M} \chi(\symp{v}{m})\, w(m)
\end{equation}
with $\Tr[\Pi(M,v)] = \frac{d^n}{|M|}$. All states in the subspace which $\Pi(M,v)$ projects onto therefore satisfy \eqref{eq:stab_def}, meaning that they form a code space. We can therefore identify this case as being a stabilizer code since it satisfies the definition in section \ref{sec:stab}. Even though finding stabilizer codes therefore just amounts to making a choice for $M$ and $v$, it does not ensure that the resulting code is good in the sense that its Hamming distance might be small or does not scale well.

\subsection{Entanglement Entropy of Stabilizer States}
\label{sec:stab_entropy}

Thanks to the structure of the symplectic product \eqref{eq:symp_prod} and the multi-particle Weyl operators defined in \eqref{eq:weyl_multi}, one can easily take the partial trace of \eqref{eq:stab_state} over a desired subsystem $B$ by writing $v = v_A \oplus v_B$ (same for $m$) and $w(m) = w(m_A) \otimes w(m_B)$ for all $m \in M$ and applying \eqref{eq:weyl_trace} to the latter term in the tensor product. The resulting reduced state is then
\begin{align}
\begin{split}
\label{eq:stab_reduced}
    \rho_A &= \Tr_B[\rho] \\
    &= \frac{d^{n_B}}{d^n} \sum_{m_A \in M_A} \chi(\symp{v_A}{m_A})\, w(m_A) \\
    &\equiv \frac{|M_A|}{d^{n_A}} \, \Pi(M_A, v_A),
\end{split}
\end{align}
where we made use of the fact that $n = n_A + n_B$ and identified \eqref{eq:stab_proj}, but this time in terms of $v_A$ and
\begin{equation}
\label{eq:M_reduced}
    M_A = \{ m_A \,|\, m_A \oplus 0_B \in M\}.
\end{equation}
This is possible since the definition of $M_A$ ensures that it is again a group (although not necessarily maximally isotropic)\footnote{Naively computing $M_A$ using \eqref{eq:M_reduced} is not efficient as such an algorithm would have $\mathcal{O}(d^n)$ runtime. A runtime that is polynomial in the system size can be achieved by instead permuting the sites that are to be traced out to the front the stabilizer matrix and then computing its reduced row echolon form. The basis vectors $b = b_A \oplus b_B$ for which $b_B \neq 0$ are then removed and for the remaining elements only $b_A$ is being considered.}.

The fact that even after tracing out a subsystem the resulting reduced state is still proportional to a projection operator makes computing the entanglement entropy straightforward. While it is possible to just directly evaluate the Von-Neumann entropy $S(A) = \Tr[\rho_A \log_d(\rho_A)]$, a more elegant and insightful approach can be made by instead considering the \emph{R\'enyi entropies}
\begin{equation}
    S^{(n)}(A) = \frac{1}{1 - n} \log_d \Tr[\rho_A^n],
\end{equation}
which have the property that
\begin{equation}
    \log_d(d^{n_A}) = S^{(0)}(A) \geq S(A) \geq S^{(2)}(A) \geq \ldots 
\end{equation}
where $S(A) = S^{(1)}(A) = \lim_{n \rightarrow 0} S^{(n)}(A)$ reproduces the ordinary Von-Neumann entropy. What makes the R\'enyi entropies interesting here is that they satisfy $S^{(n)}(A) = \log_d (\rank\rho_A)$ for all $n>0$ if the state being considered has a flat entanglement spectrum i.e.\ it is proportional to a projection operator\footnote{The proof is straightforward: Let $\rho_A = \alpha \cdot \Pi_A$, then $S^{(n)}(A) = \frac{1}{1 - n} \log_d \Tr[(\alpha \cdot \Pi_A))^n] = \frac{1}{1 - n} \log_d (\alpha^n \cdot \Tr[\Pi_A]) = \frac{1}{1 - n} \log_d (\alpha^n \cdot \rank\rho_A)$. Since $\alpha = (\rank \rho_A)^{-1}$ because of normalization we have $S^{(n)}(A) = \frac{1}{1 - n} \log_d (\rank\rho_A)^{1-n} = \log_d (\rank\rho_A)$.}. Since this is the case for the reduced stabilizer state we can use the fact that $\rank \rho_A = \frac{d^{n_A}}{|M_A|}$ to show that
\begin{equation}
    S(A) = n_A - \log_d |M_A|.
\end{equation}
If the number of basis vectors $k_A = \log_d |M_A|$ is known, then computing $S(A) = n_A - k_A$ is straightforward and numerically stable\footnote{As a sanity check, note that if $\rho_A$ is pure and therefore has $S(A) = 0$ it implies that $k_A = n_A$, which is the requirement for $M_A$ to be a maximally isotropic subspace and thus to define a (pure) stabilizer state.}.